\documentclass{article}
 \usepackage{amsmath}
 \usepackage{authblk}
\usepackage{graphicx}
\usepackage{epstopdf}

\newcommand{\md}{\ensuremath{\mathrm{d}}}       
\newcommand{\half}{\ensuremath{^1\!\!/\!_2}}    

\begin{document}

\title{Relaxation volumes of microscopic and mesoscopic irradiation-induced defects in tungsten}

\date{\today}

\author[1]{Daniel R. Mason \thanks{daniel.mason@ukaea.uk}}
\affil[1]{CCFE, UK Atomic Energy Authority, Culham Science Centre, Oxfordshire OX14 3DB, UK}

\author[1,2]{Duc Nguyen-Manh}
\affil[2]{Department of Materials, University of Oxford, Parks Road, Oxford OX1 3PH, UK}

\author[3]{Mihai-Cosmin Marinica}
\affil[3]{DEN-Service de Recherches de Metallurgie 
Physique, CEA, Universite Paris-Saclay, F-91191,
Gif-sur-Yvette, France}

\author[4]{Rebecca Alexander}
\affil[4]{Metallurgie Physique 
et Genie des Materiaux, UMET, CNRS, UMR 8207, Universite de Lille, F-59000, Lille, France }

\author[1,2]{Sergei L. Dudarev}

\maketitle

\begin{abstract}
The low energy structures of irradiation-induced defects have been studied in detail, as these determine the available modes by which a defect can diffuse or relax. As a result, there are many studies concerning the relative energies of possible defect structures, and empirical potentials are commonly fitted to or evaluated with respect to these energies.
But recently [Dudarev {\it et. al.} Nuclear Fusion 2018], we have shown how to determine the stresses, strains and swelling of reactor components under irradiation from the elastic properties of ensembles of irradiation-induced defects. These elastic properties have received comparatively little attention. Here we evaluate relaxation volumes of irradiation-induced defects in tungsten computed with empirical potentials, and compare to density functional theory results where available.
Different empirical potentials give different results, but some potential-independent trends in relaxation volumes can be identified.
We show that the relaxation volume of small defects can be predicted to within 10\% from their point-defect count.
For larger defects we provide empirical fits for the relaxation volume of as a function of size.
We demonstrate that the relaxation volume associated with a single primary-damage cascade can be estimated from the primary knock-on atom (PKA) energy.
We conclude that while annihilation of vacancy- and interstitial- character defects will invariably reduce the total relaxation volume of the cascade debris, empirical potentials disagree whether coalescence of defects will reduce or increase the total relaxation volume.
\end{abstract}


\section{Introduction}

Just as it has been long acknowledged that the effect of radiation on materials is inherently multi-scale both in time- and spatial- dimension, so it is accepted that to model these effects requires transfer of high quality data from one model to the next \cite{Marian_NuclFus2016}. The form of the data required by a coarse-grained model will vary according to its requirements, but a typical workflow in nuclear materials modelling is to find high quality structural information about individual defects from Density Functional Theory (DFT) \cite{NguyenManh_JMS2012,Dudarev_ARMR2013}, information about the cascade generation process from Molecular Dynamics (MD) \cite{Calder_JNM1993,Stoller_JNM1999,Sand_EPL2013}, and about cascade evolution using object or atomistic Kinetic Monte Carlo (KMC) \cite{Beland_PRE2011,Xu_JPCM2012,Mason_JPCM2014} or Cluster Dynamics (CD) \cite{Marian_JNM2011,Dunn_JNM2013}.
This has proved successful for modelling the experimentally observed size and distribution of irradiation-induced defects formed in pure single crystalline materials\cite{Yi_EPL2015,Mason_JPCM2014}.

Recently we have shown that it is possible to compute stresses and strains in reactor components on the macroscopic scale of centimetres and metres from the distribution of irradiation-induced defects \cite{Dudarev_NF2018}.
As a source term, this model requires the spatially varying density of relaxation volumes of defects, and so allows the direct simulation of volumetric radiation-induced swelling and the associated stresses from an atomistic or object-based model. At the nanoscale, lattice swelling is experimentally measurable using Micro-Laue diffraction \cite{Hofmann_MatLett2010}. On a macroscopic scale, predicting the stress state of components arising from irradiation is fundamental to the successful engineering design of a nuclear fission or fusion power plant \cite{Zinkle_FusEngDes2000,Bolt_JNM2002,Zinkle_ActaMat2013}.
One outstanding issue, which this paper is intended to address, is to have good data for the relaxation volumes of lattice defects, as while accessible from simulation for years, reporting this information has been somewhat neglected in favour of establishing accurate values of formation energies and ground-state structures. For example, a comprehensive compilation of data on relaxation volumes of individual self-interstitial and vacancy point defects derived from DFT calculations has been reported only recently \cite{Dudarev_PRM2018,Dudarev_NF2018,Ma_PRM2018}. 

Here we focus our attention on a single nuclear material, tungsten.
Tungsten has been chosen as a divertor material for ITER \cite{Rieth_JNM2011,Rieth_JNM2013,Bolt_JNM2002} as it has a high melting point, high thermal conductivity and high resistance to sputtering.
For our purposes tungsten is also well-suited to this preliminary study of relaxation volumes as it is nearly elastically isotropic \cite{Featherston_PhysRev1963}. This simplifies the expressions needed and so permits a simple analysis, but is by no means a requirement of the atomistic techniques used here \cite{Dudarev_PRM2018}. In section \ref{SmallClusters}, we compute the relaxation volumes of small defect clusters, and compare the results obtained with several embedded atom (EAM) potentials with those derived from DFT. In section \ref{LowEnergyClusters}, we move on to larger lattice defect objects. As the number of configurational degrees of freedom becomes very large, we focus on a standard set of idealised dislocation loops and voids, which often form a basis set for object kinetic Monte Carlo or Cluster Dynamics simulations. In section \ref{Cascades} we consider interacting groups of defect clusters generated in high energy collisions simulated by MD.

Early estimates of formation volumes for point defects in tungsten were established by Johnson\cite{Johnson_PRB1983}, using an empirical potential. This work found a negative relaxation volume for a vacancy ($-0.21\Omega _0$, where $\Omega _0$ is the atomic volume), and a positive relaxation volume for the interstitial ($+1.13\Omega _0$). DFT calculations by Kato {\it et al.} \cite{Kato_JNM2011} confirmed the relaxation volume for the vacancy as negative, at $-0.34\Omega _0$, a figure which has since been reproduced several times. The relaxation volume of a $\half\langle111\rangle$ interstitial defect was shown to be large and positive in DFT calculations at $+1.68\Omega _0$ \cite{Hofmann_Acta2015} in a small 4x4x4 supercell, later confirmed by other DFT calculations \cite{Dudarev_PRM2018,Ma_PRM2018}. 

We demonstrate that empirical potentials give varying results for the relaxation volumes of irradiation defects. This is an expected result, as these properties of defects were never originally used as input data during the parameterization of potentials. The relaxation volumes do, however, show systematic trends across potentials. It is beyond the scope of this paper to provide a comprehensive comparison of empirical potentials, instead our comparison will focus on three empirical potentials, which should give an indication of the possible variation.
\begin{itemize}
	\item
The Derlet-Nguyen-Manh-Dudarev (DND) potential \cite{Derlet_PRB2007}, has been shown to produce cascade structures that are a good match to experiment \cite{Sand_EPL2013,Yi_EPL2015,Sand_MRL2017,Mason_ActaMat2018,Mason_EPL2018}.
	\item
One of the four potential parameterizations developed by Marinica {\it et al.} (CEA-4)\cite{Marinica_JPCM2013}, which has a good balance between the predicted point- and extended defect properties. This potential was developed from the DND and AM04\cite{Ackland_JPCM2004} potentials with additional fitting to the forces on atoms in disordered systems.
	\item
A new potential parameterization by some of the authors (MNB) \cite{Mason_JPCM2017}, which is a development of the smooth and highly-transferable Ackland-Thetford potential \cite{Finnis_PMA1984,Ackland_PMA1987}, corrected to give better properties for vacancy-type structures.
\end{itemize}

We present simple empirical formulae for the relaxation volumes of defects that might be used for predicting stresses and strains in engineering components containing these defects \cite{Dudarev_NF2018}.  As tungsten is (nearly) elastically isotropic, it suffices to present results in terms of a single relaxation volume, and the relaxation volume anisotropy parameter, defined as the ratio of the smallest to largest partial relaxation volumes. We present the simple formulae required to switch between this representation and the full dipole tensor for the defect.

Finally in section \ref{comparison_linear_elasticity} we compare our results with analytical formulae derived using linear elasticity and surface energies, and in section \ref{discussion} we discuss the predictive power of our results.

\section{Relaxation volume of defect structures}

        Relaxation volumes can be computed for isolated defects and relaxed cascade configurations using several methods.
        As a validating convergence study, we compare three methods: the stress method, the cell relaxation method, and the free surfaces method: 
        \begin{itemize}
            \item The stress method. The atom positions are relaxed in a periodic supercell, with the vectors defining the supercell repeat fixed.
            The stress is computed on each atom and summed to give a single tensor for the cell. 
            The relaxation of a body free from surface tractions due to the defect is predicted from this stress using linear elasticity theory.
            The lattice vectors of the simulation cell never need to be updated.
            \item The cell relaxation method. As with the stress method, the atoms are relaxed in a periodic supercell, and the stress is computed. 
            From this a strain is computed, but in contrast to the stress method this is then applied to the supercell. 
            The vectors defining the supercell repeat are updated and the relaxation process is repeated until convergence.
            This iterative process of relaxing first the atoms and then the cell differentiates this method from the stress method.
            \item The free surface method. A large sphere of atoms is constructed and relaxed, producing a body with explicitly free surfaces. Then the defect is constructed inside the sphere and the entire structure is relaxed again.
            The volume of a (distorted) spheroid after relaxation is more difficult to compute than with a periodic supercell, as it is not clear where the surface should be drawn.
            However, the volume enclosed by the convex hull of the atoms $V_{\mathrm{hull}}$ is easy to compute using qhull \cite{qhull}.
            From this we can estimate that the volume of the sphere is $V_{\mathrm{spheroid}} = V_{\mathrm{hull}} ( R + r )^3/R^3$, where $R$ is the maximum radius of atom positions on the convex hull, and $r=\mathrm{a}_0/4$ is one quarter the lattice parameter, which is half the distance between $\{100\}$ planes in a bcc crystal.
            Note that this same excess linear correction $r$ would be required to find the volume of a periodic supercell from the convex hull of the atoms contained within.
        \end{itemize}

    In this work we do not consider the method of Kanzaki forces \cite{Kanzaki_JPCS1957,Domain_PRB2001}, or the method of matching displacements \cite{Barnett_PSS1972,Varvenne_PRB2017} for estimating relaxation volumes using the harmonic region of the crystal only.

    
    To compute the stress and hence the strain in an atomistic simulation with periodic boundary conditions we compute the dipole tensor as the integrated stress over the cell \cite{Leibfried1978,Varvenne_PRB2013}
    	\begin{equation}
        	P_{ij} = - \int_V \sigma_{ij}\left(\mathbf{r}\right) \md^3 \mathbf{r}.
    	\end{equation}
	The dipole tensor may also be expressed in terms of a symmetric dual tensor, $\Omega _{kl}$, characterizing the volumetric relaxation of the defect \cite{Dudarev_PRM2018,Dudarev_NF2018}
    	\begin{equation}
    	    \label{eqn:dualTensorDef}
        	P_{ij} \equiv C_{ijkl} \Omega_{kl},
        \end{equation}
	where $C_{ijkl}$ are the elements of the fourth-rank tensor of elastic constants. 
    From this dual tensor we can find the relaxation volume $\Omega _{rel}$ of the defect characterizing the volumetric relaxation of an elastic body free from surface tractions \cite{Leibfried1978},
    	\begin{equation}
    	    \label{eqn:relaxationVolumeDef}
        	\Omega_{rel} = \mathrm{Tr}\,\Omega \equiv \sum_{i=1}^3 \Omega ^{(i)},
		\end{equation}            
    where $\Omega ^{(i)}$ are the three eigenvalues of the tensor $\Omega _{kl}$, corresponding to the three partial relaxation volumes.	Hence we can find the elastic relaxation volume of the defect using a constant-volume calculation, if the elastic constants and dipole tensor of the defect are computed.

        For an empirical potential, $P_{ij}$ can be computed simply and analytically.
        For the embedded-atom form we compute the energy as a sum over pairwise and many-body contributions: $E = \sum_a V_a + F\left[ \rho_a \right]$, where $V_a = {^1\!/\!_2} \sum_b V( r_{ab} )$ is a pairwise interaction, $\rho_a = \sum_b \phi( r_{ab} )$ models the embedding electron density and $F\left[ \rho \right]$ is the many-body embedding energy.
        The dipole tensor is
        \begin{eqnarray}
            P_{ij}  &=& - \sum_{a,b} \left( \frac{1}{2} \left. \frac{ \partial V}{\partial r} \right|_{r_{ab}}\!+\left. \frac{ \partial F}{\partial \rho} \right|_{\rho_a} \! \left. \frac{ \partial \phi}{\partial r} \right|_{r_{ab}} \right)
                \frac{ r_{ab,i} r_{ab,j} }{r_{ab}}, \nonumber\\
			\label{analyticDipoleTensorEAM}                	
        \end{eqnarray}
        where $r_{ab}$ is the separation between atoms $a$ and $b$ and $r_{ab,i}$ is its $i^{th}$ Cartesian coordinate. 
        This is a simple sum over atoms and their neighbours, using the same first derivatives as a force calculation, and so is trivially implemented in any MD or atomistic MS code.
        The (fourth-rank) elastic constant tensor can be computed analytically from the second derivative of energy with respect to strain \cite{Finnis_PMA1984}.
        The computed elastic constants are given in table \ref{tab:elastic_constants}.
        
\begin{figure}[h!tb!]
        \centering
        \begin{minipage}[ht!]{0.45\textwidth}
            \centering
            \includegraphics[width=0.95\linewidth]{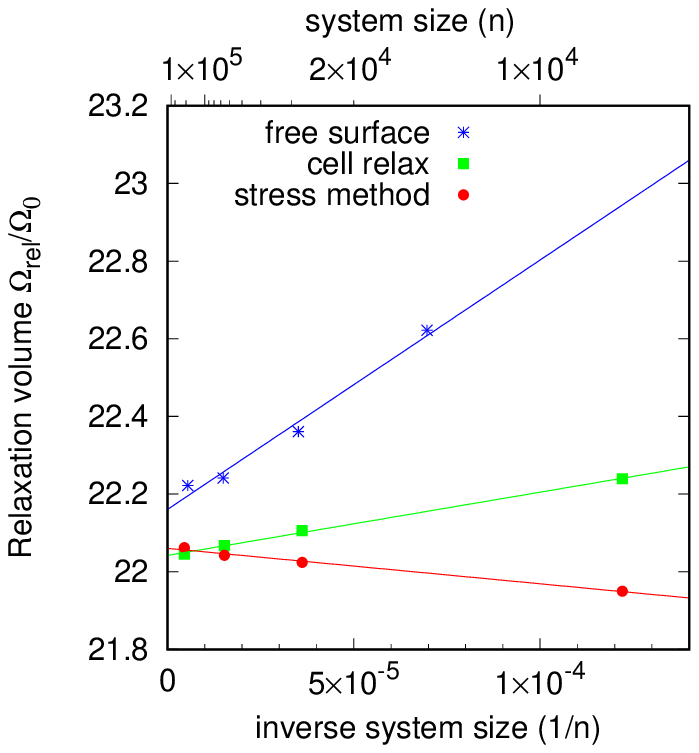}
        \end{minipage}
        \begin{minipage}[ht!]{0.45\textwidth}
            \centering
            \includegraphics[width=0.95\linewidth]{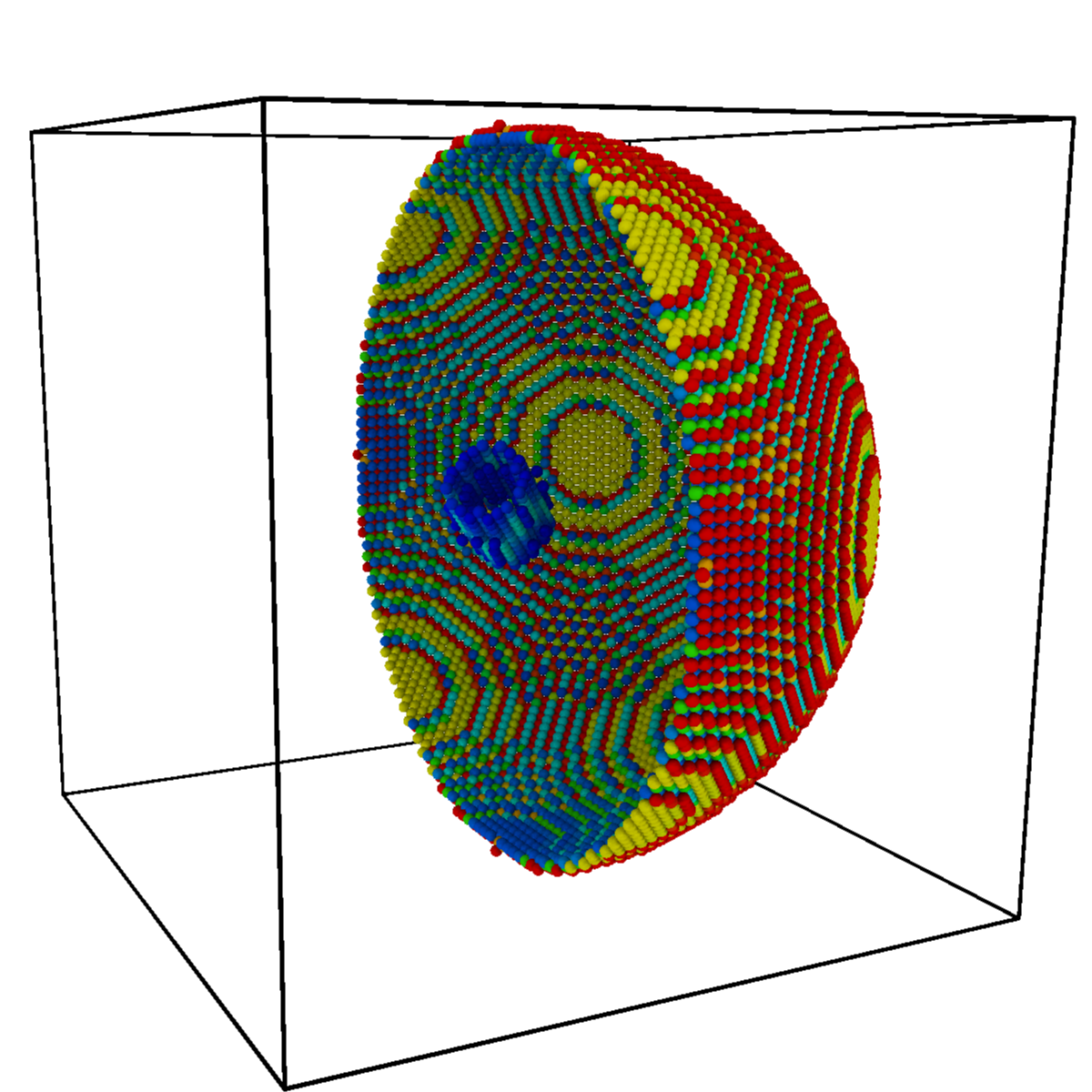}
        \end{minipage}
        \caption{
        Above: Example convergence study for relaxation volume of the 19 interstitial cluster computed using three methods.
        All three methods converge to very similar values, with a system size error inversely proportional to system size $n$.
        Best fit lines are indicated to guide the eye.
        Calculations were performed with the MNB potential. Other potentials and other defects show qualitatively similar results.
        Below: Geometry used for free-surface calculations. 
        An unsupported free sphere of atoms was constructed and relaxed, then a defect generated in the centre and the atoms relaxed again.
        The volume before and after the defect was placed was computed using qhull \cite{qhull}.
        One half of the sphere is shown \cite{Ovito}, together with high energy atoms of the defect. Atoms are coloured by excess potential energy from 0 eV (blue) to 2 eV (red). Atoms with energy under 0.1 eV are not shown.
        In this image a 55 interstitial loop is embedded in a sphere of 180000 atoms, and relaxed with the MNB potential.
        } 
        \label{fig:convergence_study}    
        \end{figure}

        A plot showing the convergence of the numerical procedure for computing the relaxation volume of a 19-interstitial loop with system size is shown in figure \ref{fig:convergence_study}.
        Extrapolating to infinite system size suggests all three methods converge to the same result, though at any given finite system size there will be an error, typically scaling as inverse system size $1/n$ ( voids relaxed with explicitly free surfaces being an exception, converging as the inverse radius of the free sphere $1/R$. )
        This $1/n$ convergence was also observed recently by Varvenne \& Clouet \cite{Varvenne_PRB2017}, who attributed this leading error term to the interaction between periodic images.
        Some indicative data are also given in table \ref{tab:convergence_study} proving that the stress method with large supercells is suitable for the structures considered in this work.
        Relaxation volumes computed with the EAM potentials in this paper are computed using the dipole tensor method at a converged supercell size.
        Relaxation volumes computed with DFT were computed using the full cell relaxation method.
        
\begin{table*}
\centering
\begin{tabular}{   l | ccc  }
          &   cell relaxation &   stress method &   free surfaces   \\
\hline
1v                          &   -0.367  &   -0.368	&   -0.350   \\
15v$_{sph}$                 &      -3.599	&   -3.625	&   -3.731  \\
65v$_{sph}$                 &     -9.658	&   -9.760	&   -10.351  \\
259v$_{sph}$                &     -21.634	&   -21.806	&   -24.291  \\
\hline
1i$_{\langle111\rangle}$    &     1.399   &	1.400   &  	1.415  \\
19i$_{\langle111\rangle}$   &     22.045  &	22.062  &  	22.222  \\
55i$_{\langle111\rangle}$    &     60.325  &	60.250  &  	60.705  \\
199i$_{\langle111\rangle}$  &   209.946  & 209.283   &	210.947
\end{tabular}
\caption{
Relaxation volumes for vacancy defects ( monovacancy, and 15,65,259 vacancy spheres ) and interstitial defects ( $1/2\langle 111\rangle$ dumbbell, 19,55,199 $\langle 111\rangle$ interstitial loops ).
The volumes are expressed in units of atomic volume, computed using the MNB potential.
The cell relaxation and stress method calculations used a $48\times 48\times 48$ unit cell supercell.
The free surfaces method used a sphere with an initial radius of $28$ unit cells.
It is clear that the stress method does indeed reproduce the relaxation volume of the defect in a body with traction free surfaces at large system size, and so is an acceptable faster alternative to the full cell relaxation calculation.
}
\label{tab:convergence_study}
\end{table*}

\begingroup
\begin{table*}
\centering
\begin{tabular}{   l | c | rrr | rrrr | c }
method 	&	$\mathrm{a}_0$		&	$c_{11}$    &   $c_{12}$    &   $c_{44}$	&	$\gamma_{\langle 110\rangle}$   &   $\gamma_{\langle 100\rangle}$ & $\gamma_{\langle 211\rangle}$ & $\gamma_{\langle 111 \rangle}$ & $\gamma$	\\
\hline
DFT		&   3.186       &   3.229   &   1.224   &   0.888       &     0.200 &   0.245   &   0.215   &   0.219    &  0.229    \\ 
MNB		&   3.1652	    &   3.222   &   1.263   &   0.998	   &     0.218 &   0.239   &   0.241   &   0.257    &  0.234    \\
DND		&   3.1652      &   3.3881  &   1.304   &   1.031       &     0.150 &   0.187   &   0.185   &   0.161    &  0.174    \\
CEA4	&   3.14339     &   3.265  &    1.262   &   1.004    	&     0.156 &   0.183   &   0.200   &   0.186    &  0.190    \\
expt    &   3.1652      &   3.324   &   1.279   &   1.018       &           &           &           &           &   0.229   \\
\end{tabular}
\caption{
The lattice constant $\mathrm{a}_0$ (in \AA) and elastic constants (in eV/\AA$^3$) of bcc tungsten, from \cite{Hofmann_Acta2015}.
The surface energies of bcc tungsten in eV/\AA$^2$.
Average surface energy computed using equation \ref{eqn:average_surface_energy}.
Experimental elastic constants from ref \cite{Featherstone_PR1963}, and surface energy from ref \cite{deBoer1988}.
}
\label{tab:elastic_constants}
\end{table*}
\endgroup


\subsection{Small defect structures}
\label{SmallClusters}

In this section we compute the elastic properties of small defect clusters.
For sufficiently small clusters we can perform a fairly comprehensive survey of possible structures, and so find the bounds of the variation of the relaxation volumes. 
The question of which are the most significant set of small clusters to use is rather more difficult. 
Generally, for a cluster containing $N$ point defects, we might expect to find only a small number of structures within a few meV of the ground state. 
At low temperatures these are the only ones which need to be considered in equilibrium.
But radiation damage is an inherently non-equilibrium process. 
The system can generally reduce its internal energy by coalescing clusters, and the true equilibrium is only found when nearly all defects have recombined or diffused to sinks.
The structures that may actually be found at some time after the cooling of a displacement cascade could therefore be, briefly at least, rather exotic.

\begin{figure}[htb!]
\centering
\includegraphics[width=7.5cm]{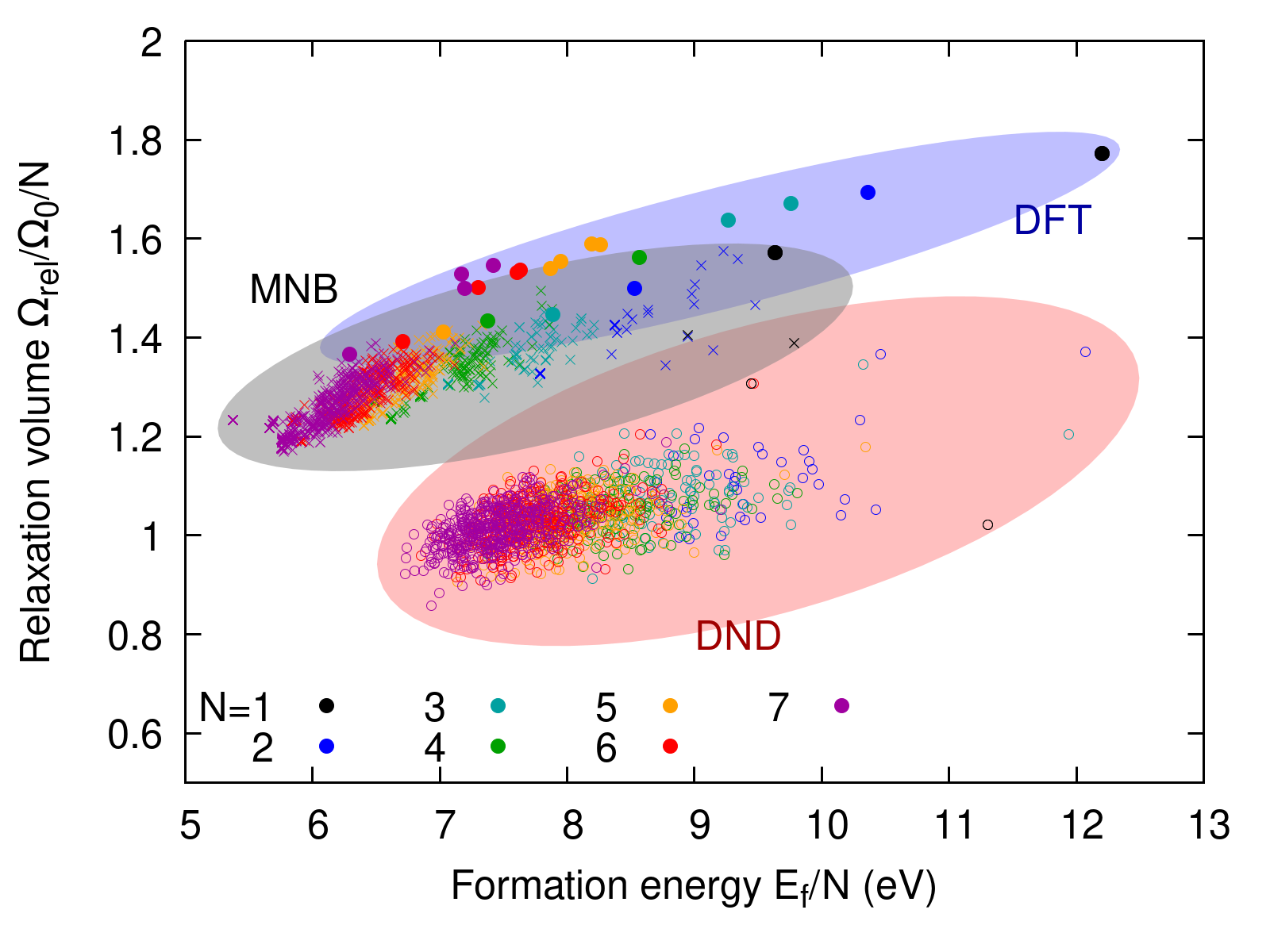}\\
\caption{
Relaxation volumes of randomly generated interstitial defect clusters. DFT values for energies and relaxation volumes from Ref. \cite{Dudarev_NF2018} are shown with filled circles. Crosses: the values computed with MNB potential, open circles are the values computed using the DND potential.
Shaded ellipses are drawn to guide the eye to the regions covered by data generated uisng the relevant potentials.
Note that the DND potential tends to predict a higher formation energy and lower relaxation volume of a defect cluster than the MNB potential.
} 
\label{fig:random_i}       
\end{figure}

Randomly generated interstitial clusters were generated by placing $N$ additional atoms into an otherwise perfect crystal, then relaxing. 
The extra atoms were placed at random into $[ l,m,n ]\mathrm{a}_0/4$ crystal positions, with $0 \le lmn < 4$, with the constraint that an atom was not placed if another was already within a distance $\mathrm{a}_0/2$. All extra atoms were placed in a central block of $2\times 2\times 2$ unit cells.
The relaxation volumes as a function of formation energy and cluster size are presented in figure \ref{fig:random_i}.
We have not put results from CEA4 on this scatter plot. 
CEA4 has a good deal of structure in its potential, which allows for a very large number of metastable high energy defect clusters to form. 
On figure \ref{fig:random_i}, this would appear as an almost structureless cloud and we conclude our method of randomly generating interstitial clusters is not suited to this potential.

DFT calculations of small interstitial clusters were performed for this work using the VASP \textit{ab initio} simulation code, using the PAW method \cite{PhysRevB.47.558,PhysRevB.54.11169,KRESSE199615} with semi-core electrons included
through the use of pseudo-potentials. It is important to emphasize that the inclusion of semi-core electrons in the valence states has a significant effect on the predicted
formation energies of self-interstitial atom (SIA) defects for all the bcc transition metals \cite{NguyenManh_JMS2012,NguyenManh_PRB2006,NguyenManh_NIMB2015}, and play important role on the quality of inter-atomic potential in predicting non-equilibrium properties in tungsten from cascade simulations \cite{Sand_JNM2016}. Exchange-correlation effects were described using the Perdew-Burke-Ernzerhof generalised gradient approximation\cite{Perdew_PRL1996}. A kinetic energy cut-off of 400 eV was used, with a $3\times 3\times 3$ Monkhorst-Pack grid for electron density k-points employed in the case with super-cell (8x8x8) calculations ($1024+N$ atoms, with N up to 22 atoms) for the $\half \langle 111\rangle$ and $\langle 100\rangle$ interstitial defects. 
The set of interstitial defect clusters used was the same as in ref \cite{Alexander_PRB2016}, with the difference that in the earlier work the energies reported were in the constant volume approximation employing the Varvenne cell size correction\cite{Varvenne_PRB2013}, whereas here the full cell relaxation method was used.
The slab model including a 18 \AA vacuum between the top and bottom surfaces has been employed for calculations of W(100), W(110), W(111) and W(211) surface energies where 15, 15, 24 and 24 layers were used, respectively, to ensure the DFT convergence. 

\begin{figure}[htb!]
\centering
\includegraphics[width=7.5cm]{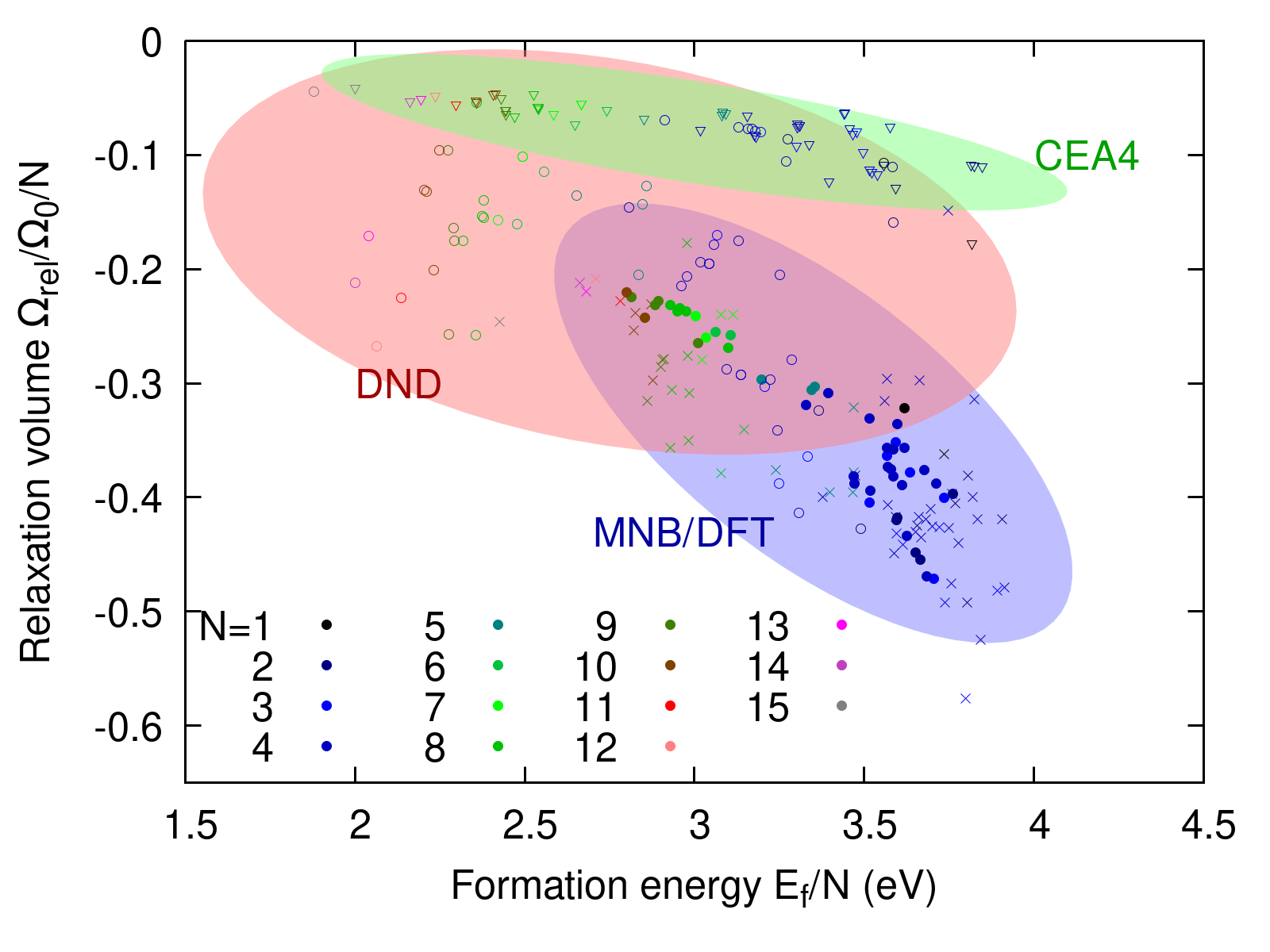}\\
\caption{
Relaxation volumes of randomly generated vacancy defect clusters.
DFT values for energies are from Ref. \cite{Mason_JPCM2017}, the relaxation volumes were computed in this study, and are shown with filled circles.
Crosses: the values computed with MNB potential, open circles with DND potential.
Shaded ellipses are drawn to guide the eye to the regions covered by the potentials.
Note that the DND potential tends to predict a lower formation energy and smaller magnitude relaxation volume than the MNB potential.
The CEA potential predicts the smallest relaxation volume.
The MNB potential data have a high degree of overlap with the DFT relaxation volume data.
} 
\label{fig:random_v}       
\end{figure}

Randomly generated vacancy clusters were produced by removing atoms on a random path through an otherwise perfect crystal. The path was allowed to move in $\langle 111\rangle$ and $\langle 100 \rangle$ directions, and allowed to overlap itself. A path of length $L$ steps leads to $\le L$ vacancies placed in a loose cluster. 
These vacancy clusters were then relaxed and the lowest energy structures were passed to DFT for a comparison calculation. It is important to emphasize that in the present DFT calculations for both SIA and vacancy clusters, the full cell relaxation method has been adopted to investigate the relaxation volume of defect structures. Details of the DFT calculations can be found in Ref. \cite{Mason_JPCM2017}. Results are shown in figure \ref{fig:random_v}.

Two trends are immediately apparent in figures \ref{fig:random_i} and \ref{fig:random_v}.
Firstly, we see that larger clusters have a lower formation energy per point defect. 
This is just an illustration of the reduction in energy due to coalescence.
But we also can see that the magnitude of the relaxation volume reduces with cluster size. By coalescing, the volume strain on the lattice required to accommodate defects is reduced.

The second trend we see is that the distribution of relaxation volumes is small, of order $\pm 0.1$ atomic volume per point defect.
A visually striking consequence of this is that the scatter plots for different potentials do not significantly overlap, but this is of limited physical significance as the offset for each potential is a consequence of its fitting, and the fitting did not consider relaxation volume.

The true physical significance of this result is that even if the exact structure of a defect cluster is not known, its relaxation volume can be estimated from its size alone. The accuracy in the estimate of the relaxation volume may be rather low, as the different potentials given different offsets, but this lack of accuracy will itself be small compared to the current accuracy in modelling the time-dependent evolution of cascades.

Tabulated values of the  relaxation volumes of low energy small defect clusters are given in table \ref{tab:volume_lowest}.

\begin{table*}
\centering
\begin{tabular}{   l | l | rrrrrrr   }
method 	&	&   \multicolumn{7}{c}{$\Omega_{rel}/\Omega_0$}   \\
        & character  &   N=1     &   2       &   3       &   4       &   5   &    6   &   7 \\
\hline
DFT		& vac   &	 -0.32  &   -0.84   &   -1.21   &   -1.28   &   -1.48  &    -1.53   &   -1.69             \\
MNB		& vac   &    -0.36  &   -0.85   &   -1.29   &   -1.60   &   -1.88  &    -2.27   &   -1.96              \\
DND		& vac   &    -0.11  &   -0.17   &   -0.51   &   -0.58   &   -0.68  &    -0.96   &   -1.10      \\
CEA4	& vac   &    -0.18  &   -0.22   &   -0.28   &   -0.32   &   -0.35  &    -0.45   &   -0.46        \\
\hline
DFT		& i$\langle 111 \rangle$   &	 1.57   &   3.00   &   4.34      & 5.73      & 7.06      &   8.35    &   9.56            \\
MNB		& i$\langle 111 \rangle$   &     1.40   &   2.65   &    3.92   &    5.08   &    6.25   &    7.39   &       8.63     \\
DND		& i$\langle 111 \rangle$   &     1.31  &    2.41   &   3.48   &   3.85   &   4.53  &    5.76   &   6.45     \\
CEA4	& i$\langle 111 \rangle$   &     1.25  &    2.38   &   4.57   &   4.80   &   6.06  &    7.36   &   6.97       \\
\hline
DFT		& i$\langle 100 \rangle$   &	 1.77  &   3.39   &   4.91   &   6.25   &   7.77  &    9.00   &   10.70            \\
DFT		& i C15                    &	       &   3.92   &          &   6.59   &   8.19   &   9.37  &    11.00              \\
\end{tabular}
\caption{
The relaxation volumes ($\Omega_{rel}/\Omega_0$) of the lowest energy vacancy and interstitial clusters $1\le N \le 7$.
These relaxation volumes were computed using full relaxation of the simulation cell and atom positions.
}
\label{tab:volume_lowest}
\end{table*}

\subsection{Lowest energy defect structures}
\label{LowEnergyClusters}

Having considered randomly-generated defect clusters, we now turn our attention to larger low energy defects. Experimentally, both the $\langle 100 \rangle$ and $\half \langle 111 \rangle$ interstitial and vacancy-type loops are observed in ion-irradiated ultra-high purity tungsten foil in the TEM\cite{Yi_PM_2013,Mason_JPCM2014}.
We will consider these four loop types as idealised, planar, circular, prismatic loops, such as might be the basis set for time-evolution in object kMC or CD.
To these objects we add the spherical voids. It may be that, especially at larger sizes, the facetted voids, or hexagonal prismatic loops have slightly lower energy \cite{fikar_2017,Fikar_NME2018}. 
For our purposes it is not necessary to guarantee that we have the true ground state of a defect cluster as we are attempting to find trends that govern the variation in relaxation volume, independent to the choice of interatomic interaction potential.

\begin{figure}[h!tb!]
\centering
\includegraphics[width=7.5cm]{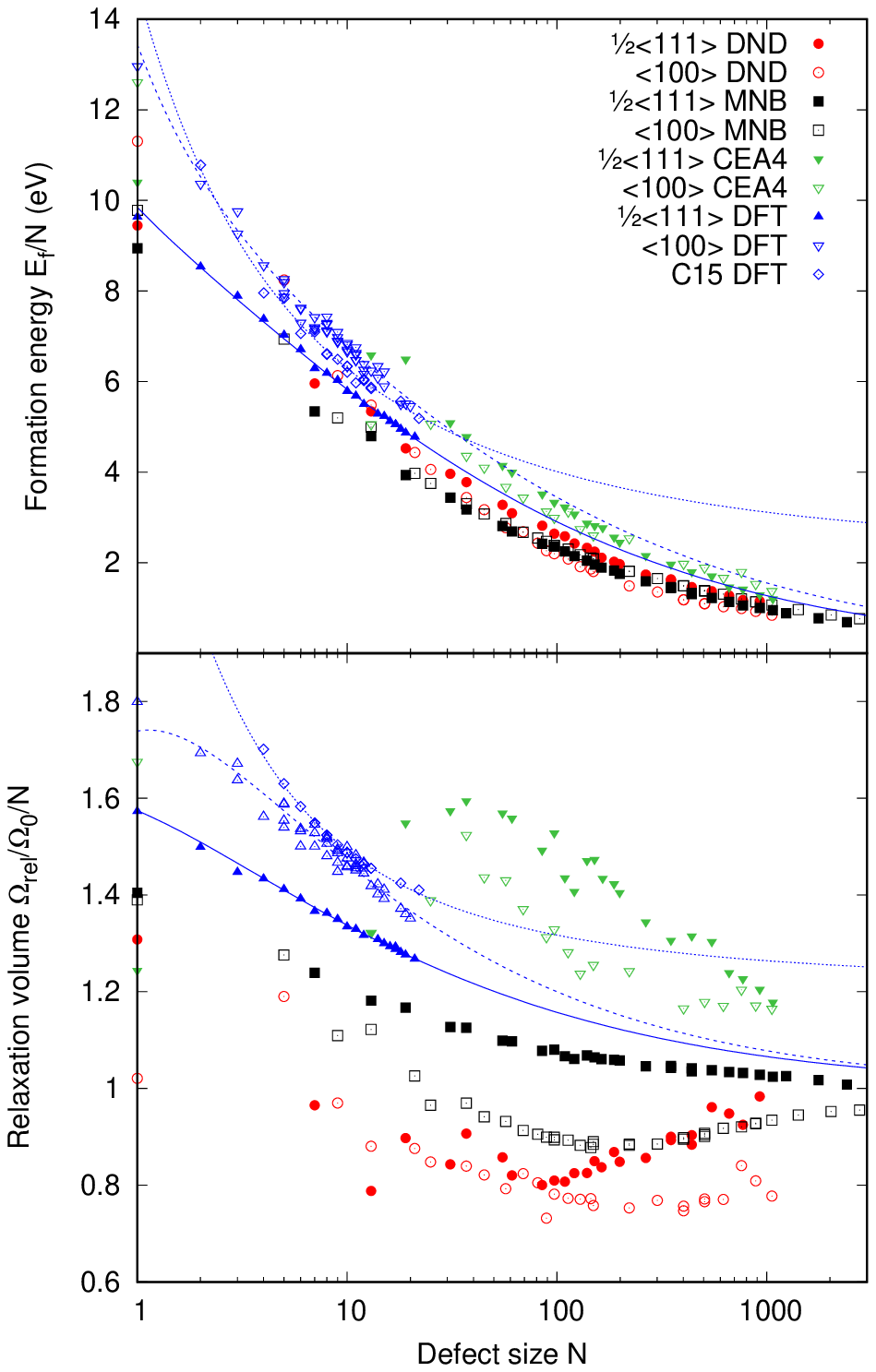}\\
\caption{
Formation energy and relaxation volume of low energy interstitial defect clusters.
All clusters and loops are of circular shape.
DFT values for formation energies are extrapolated with lines fitted to $E_f = a_0 \sqrt{N} \ln{N} + a_1 \sqrt{N} + a_2$ (see table \ref{tab:regression_energy_loop}).
DFT values for relaxation volumes  are extrapolated with lines fitted to $\left| \Omega _{rel}/\Omega _0 \right| = N + b_0 \sqrt{N} \ln{N} + b_1 \sqrt{N} + b_2$ ( see table \ref{tab:regression_volume_loop} ).
Note that the energies for ideal interstitial defects computed with the potentials are very similar, but the relaxation volumes differ considerably, with the DND potential typically predicting smaller values and CEA4 larger.
} 
\label{fig:perfect_i}       
\end{figure}

\begin{figure}[h!tb!]
\centering
\includegraphics[width=7.5cm]{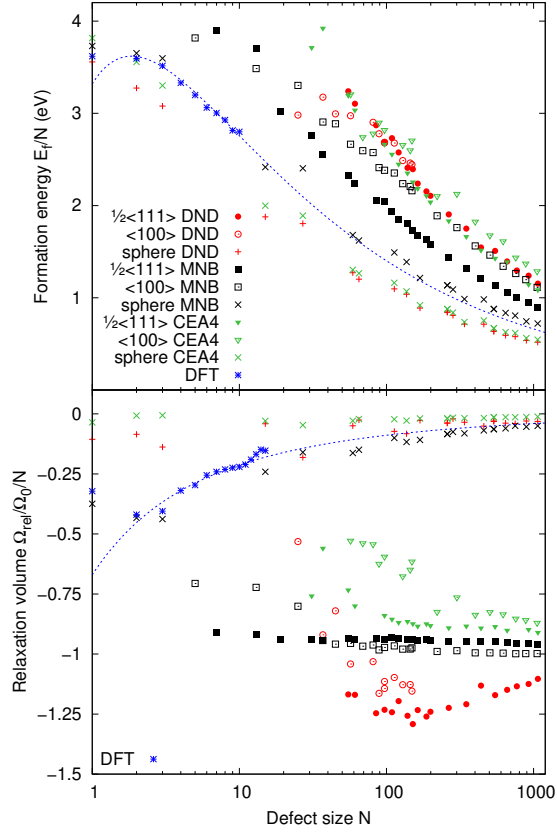}\\
\caption{
Formation energy and relaxation volumes of low energy vacancy defect clusters.
The loops were generated with circular shapes, and the voids as spheres.
DFT computed energies of the low-energy vacancy clusters from Ref \cite{Mason_JPCM2017}, with an extrapolated line fitted to $E_f = a_0 N^{2/3} + a_1$. 
DFT computed relaxation volumes for the same structures in Ref \cite{Mason_JPCM2017} fitted to $\Omega _{rel}/\Omega_0 = b_0 N^{2/3} + b_1$.
} 
\label{fig:perfect_v}       
\end{figure}

We construct the prismatic loops and spherical voids using the procedure proposed in \cite{Gilbert_jpcm2008}. Formation energies and relaxation volumes for the interstitial defects are presented in figure \ref{fig:perfect_i}, and for vacancy defects in figure \ref{fig:perfect_v}. The formation energies are included here as a comparison to previous studies, and we find that our answers agree with results given in Refs. \cite{Gilbert_jpcm2008,Fikar_NME2018} to order of the symbol size.

We can see from figure \ref{fig:perfect_i} that the relaxation volume of a large interstitial loop tends to $\lim_{N\rightarrow \infty} \Omega_{rel} / \Omega _0 = N$, where $\Omega _0$ is the atomic volume. 
This result confirms that the volume per atom in an edge dislocation, which is a semi-infinite plane of atoms embedded in a crystal lattice, must be $\Omega _0$. 
What is more surprising is the slow rate at which the result converges to this answer. Though the vertical scale in figure \ref{fig:perfect_i} is chosen to exaggerate the effect, nevertheless it can be that loops need to be well over one hundred point defects, perhaps even over one thousand before this limit can truly be said to be reached. 
To give the reader an idea of the spatial scale involved, a circular $\half \langle 111 \rangle$ dislocation loop containing one thousand interstitials has the diameter of 8.5nm. A second interesting feature of figure \ref{fig:perfect_i} is that the relaxation volume is not necessarily a monotonic function of $N$. 

A regression analysis of the relaxation volume as a function of size for interstitial loops suggests that an excellent fit can be found for the empirical form
	\begin{equation}
		\Omega_{rel}/\Omega _0  = N + b_0 \sqrt{N} \ln{N} + b_1 \sqrt{N} + b_2.
    \end{equation}
We fit this using least squares fitting to $(\Omega_{rel}/\Omega_0 - N)/\sqrt{N}$, and find error bar estimates using bootstrapping \cite{Diaconis_SciAm1983}.
Formation energies have been fitted to $\Omega_{rel}/\Omega _0/\sqrt{N}$ using the same method.
Fits for the relaxation volumes and formation energies are given in tables \ref{tab:regression_energy_loop} and \ref{tab:regression_volume_loop} respectively.

In figure \ref{fig:perfect_v} we see that the relaxation volume of a large vacancy loop tends to $\lim_{N\rightarrow \infty} \Omega_{rel} / \Omega_0 = -N$, and again may be non-monotonic. At small cluster sizes $N < 30$, however, it is not clear if there should be a single function describing the relation between the relaxation volume and defect size $N$. Small vacancy loops are unstable with respect to their transformation to open platelets\cite{Gilbert_jpcm2008} and subsequently to spherical voids, particularly for the DND and CEA-4 potentials, so the smallest relaxed clusters may not be strictly classified as `loops'. We have omitted small vacancy clusters which show significantly different elastic properties to large loops.

A regression analysis of relaxation volume and formation energy of the C15 structures computed using DFT are tabulated in table \ref{tab:regression_energyvolume_C15}. 
The formation energy is fitted to $E_f =  a_0 N + a_1 N^{2/3} + a_2$, indicating that the energy is driven by both volume and surface energy terms.
The relaxation volume is well fitted by this same form, $\Omega_{rel}/\Omega_0 = b_0 N + b_1 N^{2/3} + b_2$.
It is not required that the relaxation volume per C15 interstitial tends to $\lim_{N\rightarrow \infty} \Omega_{rel} / \Omega_0 = N$, as the structure is not bcc.
We note that the C15 structures are higher energy than the  $\half \langle 111 \rangle$ interstitial dislocation loops.
Our fitting suggests that at larger defect sizes C15 will be less stable than dislocation loops.

A regression analysis of relaxation volume and formation energy of voids are tabulated in table \ref{tab:regression_energyvolume_void}. 
The formation energy is fitted to $E_f =  a_0 N^{2/3} + a_1 $, indicating that the energy is driven by surface energy alone.
A regression analysis of the relaxation volume of a void as a function of void size suggests that an excellent fit can be found for the form $\Omega _{rel}/\Omega _0 = b_0 N^{2/3} + b_1$, for $N>5$. The two-thirds power implies that the elastic relaxation of a void is driven by the minimization of the surface energy of the void, and the resulting elastic contraction of the material around it. 
Results for the relaxation volumes are given in table \ref{tab:regression_energyvolume_void}, and the DFT fit is shown on figure \ref{fig:perfect_v}.
The MNB potential predictions are in excellent agreement with the DFT results for the relaxation volume. The relaxation volumes of the lowest energy vacancy clusters for $1\le N \le 7$ are given in table \ref{tab:volume_lowest}.
We analyse the relaxation volume of the void using linear elasticity theory in section \ref{relax_vol_void}.

\begingroup
\begin{table*}
\centering
\begin{tabular}{   l | l | r | r | r  }
method 	&	structure				&	$a_0$		& $a_1$			&	$a_2$				\\
\hline
DFT		&   i$\langle 111 \rangle$ 	&	 $	4.838	\pm 	0.039	$ & $ 	6.183	\pm 	0.165	$ & $	3.617	\pm 	0.231	$ \\    
DFT		&	i$\langle 100 \rangle$ 	&	$	6.021	\pm 	0.032	$ & $ 	4.760	\pm 	0.127	$ & $	8.225	\pm 	0.172	$ \\    
\hline
MNB		&   i$\langle 111 \rangle$ 	&	$	3.394	\pm 	0.007	$ & $ 	7.081	\pm 	0.052	$ & $	1.803	\pm 	0.156	$ \\    
MNB		&   i$\langle 100 \rangle$ 	&	$	4.773	\pm 	0.018	$ & $ 	1.103	\pm 	0.127	$ & $	9.760	\pm 	0.358	$ \\    
MNB		&   v$\langle 111 \rangle$ 	&	$	4.155	\pm 	0.016	$ & $ 	-0.006	\pm 	0.116	$ & $	6.033	\pm 	0.376	$ \\    
MNB		&   v$\langle 100 \rangle$ 	&	$	5.588	\pm 	0.034	$ & $ 	-2.267	\pm 	0.245	$ & $	3.197	\pm 	0.837	$ \\   
\hline
DND		&   i$\langle 111 \rangle$ 	&	$	3.456	\pm 	0.025	$ & $ 	10.113	\pm 	0.166	$ & $	-0.741	\pm 	0.501	$ \\  
DND		&   i$\langle 100 \rangle$ 	&	$	1.479	\pm 	0.036	$ & $ 	15.870	\pm 	0.247	$ & $	-3.318	\pm 	0.693	$ \\  
DND		&   v$\langle 111 \rangle$ 	&	$	4.684	\pm 	0.079	$ & $ 	5.051	\pm 	0.604	$ & $	2.818	\pm 	2.489	$ \\    
DND		&   v$\langle 100 \rangle$ 	&   $	4.005	\pm 	0.280	$ & $ 	15.006	\pm 	1.780	$ & $	-64.593	\pm 	4.850	$ \\ 
\hline
CEA4	&   i$\langle 111 \rangle$ 	&   $	2.706	\pm 	0.033	$ & $ 	21.547	\pm 	0.228	$ & $	-11.305	\pm 	0.706	$ \\ 
CEA4	&   i$\langle 100 \rangle$ 	&	$	7.149	\pm 	0.080	$ & $ 	-3.331	\pm 	0.531	$ & $	16.124	\pm 	1.624	$ \\  
CEA4	&   v$\langle 111 \rangle$ 	&	$	4.675	\pm 	0.075	$ & $ 	3.077	\pm 	0.537	$ & $	15.172	\pm 	1.940	$ \\   
CEA4	&   v$\langle 100 \rangle$ 	&   $	1.993	\pm 	0.216	$ & $ 	31.441	\pm 	1.669	$ & $	-126.319	\pm 	6.691	$ \\
\end{tabular}
\caption{
Linear regression fits for the formation energy (in eV) of interstitial and vacancy loops fitted to the form $E_f = a_0 \sqrt{N} \ln{N} + a_1 \sqrt{N} + a_2$.
The DFT values shown are from this study using the cell relaxation method.
Note that the fits should only be considered accurate in the ranges covered by the points in figures \ref{fig:perfect_i}, \ref{fig:perfect_v}.
}
\label{tab:regression_energy_loop}
\end{table*}
\endgroup

\begingroup
\begin{table*}
\centering
\begin{tabular}{   l | l | r | r | r  }
method 	&	structure				&	$b_0$		&$b_1$			&	$b_2$				\\
\hline
DFT		&   i$\langle 111 \rangle$ 	&	  $    -1.983 \pm     0.007 $ & $    2.614  \pm     0.026 $ & $   -3.195  \pm     0.035   $ \\ 
DFT		&   i$\langle 100 \rangle$ 	&	  $    -1.977 \pm     0.003 $ & $    2.698  \pm     0.011 $ & $   -3.330  \pm     0.015   $ \\ 
\hline
MNB		&   i$\langle 111 \rangle$ 	&	 $    0.008  \pm     0.002 $ & $    0.738  \pm     0.010 $ & $   -0.179  \pm     0.021   $ \\ 
MNB		&   i$\langle 100 \rangle$ 	&	 $    -0.420 \pm     0.007 $ & $    0.622  \pm     0.048 $ & $   2.127   \pm     0.158   $ \\ 
MNB		&   v$\langle 111 \rangle$ 	&	 $    0.253 \pm     0.002 $ & $     -0.480 \pm     0.008 $ &                               \\  
MNB		&   v$\langle 100 \rangle$ 	&	 $    -0.155  \pm     0.003 $ & $    1.053 \pm     0.015 $ &                               \\                            
\hline
DND		&   i$\langle 111 \rangle$ 	&	 $    -0.061 \pm     0.016 $ & $    -1.566 \pm     0.096 $ & $   2.945   \pm     0.227   $ \\
DND		&   i$\langle 100 \rangle$ 	&    $    -1.348 \pm     0.006 $ & $    3.534  \pm     0.032 $ &                               \\                          
DND		&   v$\langle 111 \rangle$ 	&	 $     0.112 \pm     0.009 $ & $   -5.462  \pm     0.055 $ & $    25.068 \pm     0.255   $ \\ 
DND		&   v$\langle 100 \rangle$ 	&	 $     0.377 \pm     0.081 $ & $   -6.831  \pm     0.492 $ & $    39.429 \pm     1.217   $ \\ 
\hline
CEA4	&   i$\langle 111 \rangle$ 	&    $    -0.050 \pm     0.018 $ & $    7.440  \pm     0.116 $ & $   -22.196 \pm     0.318   $ \\
CEA4	&   i$\langle 100 \rangle$ 	&    $    0.845  \pm     0.027 $ & $    -0.935 \pm     0.170 $ & $   1.595   \pm     0.435   $ \\
CEA4	&   v$\langle 111 \rangle$ 	&	 $    0.384 \pm     0.015 $ & $    -0.186  \pm     0.101 $ & $    2.075  \pm     0.330   $ \\
CEA4	&   v$\langle 100 \rangle$  &    $    0.554 \pm     0.041 $ & $    -0.144  \pm     0.316 $ & $    13.867 \pm     1.194   $ \\
\end{tabular}
\caption{
Linear regression fits for the relaxation volume ($\Omega_{rel}/\Omega _0$) of interstitial and vacancy loops fitted to the form $\Omega_{rel}/\Omega _0  = \pm N + b_0 \sqrt{N} \ln{N} + b_1 \sqrt{N} + b_2$, where the positive and negative signs are for interstitial and vacancy loops respectively.
Note that the fits should only be considered accurate in the ranges covered by the points in figures \ref{fig:perfect_i}, \ref{fig:perfect_v}.
}
\label{tab:regression_volume_loop}
\end{table*}
\endgroup

%

\begin{table*}
\centering
\begin{tabular}{   l | rr | rr  }
method 	&   $a_0$		&	$a_1$   &	$b_0$		&	$b_1$	\\
\hline
DFT		&	$6.66 \pm 0.06$	    &	$-3.34 \pm 0.20$
        &   $-0.40 \pm 0.02$	&	$-0.27 \pm 0.07$	\\
MNB		&   $7.35 \pm 0.06$     &   $-7.4 \pm 3.0$     
        &   $-0.50 \pm 0.01$	&   $-0.77 \pm 0.40$	\\
DND		&    $5.25 \pm 0.05$    &   $-1.9 \pm 2.3$  
        &   $-0.31 \pm 0.02$    &   $0.28 \pm 1.12$   	\\
CEA4	&   $5.56 \pm 0.06$     &   $-3.2 \pm 2.8$ 
        &   $-0.122 \pm 0.003$  &   $0.10 \pm 0.12$   	\\
\end{tabular}
\caption{
Linear regression fits for the formation energy (in eV) and relaxation volume $\Omega_{rel}/\Omega_0$ of vacancy clusters and voids.
The energy is fitted to the form $E_f = a_0 N^{2/3} + a_1$, indicating a domination by surface energy.
The relaxation volume is fitted to the form $\Omega_{rel}/\Omega _0 = b_0 N^{2/3} + b_1$, indicating this too is driven by surface energy.
The DFT data use the lowest energy structures for $2\le N\le 12$, the EAM data use the lowest energy structures for $6\le N\le 10$ and spherical voids for $15 \le N \le 1067$.
}
\label{tab:regression_energyvolume_void}
\end{table*}

\begin{table*}
\centering
\begin{tabular}{   l | rrr | rrr  }
method 	&	$a_0$		&	$a_1$ &	$a_2$   & 		$b_0$		&	$b_1$ &	$b_2$     \\
\hline
DFT		&	$     2.37 \pm     0.06 $ & $     7.39 \pm     0.20 $ & $     4.92 \pm     0.29 $     
        &   $     1.22 \pm     0.01 $ & $     0.39  \pm     0.03 $ & $     0.87 \pm     0.04 $  \\
\end{tabular}
\caption{
Linear regression fits for the formation energy (in eV) and relaxation volume $\Omega_{rel}/\Omega_0$ of interstitial clusters in the C15 structure.
Energies are fitted to $E_f = a_0 N + a_1 N^{2/3} + a_2$, as suggested by ref \cite{Alexander_PRB2016}, indicating terms dependent on volume and surface area.
Relaxation volume are fitted to $\Omega_{rel}/\Omega_0 = b_0 N + b_1 N^{2/3} + b_2$.  
The structures taken were those used in ref \cite{Alexander_PRB2016}, using the cell relaxation method, with sizes $2\le N \le 22$.
}
\label{tab:regression_energyvolume_C15}
\end{table*}

\subsection{The anisotropy of the elastic relaxation}
\label{anisotropy}
In equation \ref{eqn:relaxationVolumeDef} we expressed the total relaxation volume as the sum of the three partial relaxation volumes. These three partial volumes, plus the corresponding eigenvectors, completely specify the tensor $\Omega _{kl}$, and hence the dipole tensor through eq. (\ref{eqn:dualTensorDef}).

For cubic crystals, it is straightforward to invert the matrix equations and recover the dipole tensor.
For the perfect defect cluster shapes considered in section \ref{LowEnergyClusters}, one eigenvector describes both the Burgers vector and normal of the loops, and so this task can be accomplished simply and analytically.

For a $\half[ 111 ]$ loop the dipole tensor has the general form 
    \begin{equation}
        P_{\half[ 111 ]} = \left( \begin{array}{ccc} a & b & b \\ b & a & b \\ b & b & a \end{array} \right),
    \end{equation}
where $a$ and $b$ are numerical parameters with units of energy. The dipole tensor for other symmetry related $\half\langle 111 \rangle$ loops is found by taking negative signs in the off-diagonal elements as appropriate.
In cubic symmetry, we can find the elastic compliance tensor ($S \equiv C^{-1}$).
Its representation as a matrix in Voigt notation has the simple form:
    \begin{equation}
        S = \left(  \begin{array}{cccccc} 
            \frac{ c_{11}+c_{12} }{d} & \frac{ -c_{12} }{d} & \frac{ -c_{12} }{d} & 0 & 0 & 0 \\
            \frac{ -c_{12} }{d} & \frac{ c_{11}+c_{12} }{d} & \frac{ -c_{12} }{d} & 0 & 0 & 0 \\
            \frac{ -c_{12} }{d} & \frac{ -c_{12} }{d} & \frac{ c_{11}+c_{12} }{d} & 0 & 0 & 0 \\
            0 & 0 & 0 & \frac{1}{c_{44}} & 0 & 0 \\
            0 & 0 & 0 & 0 & \frac{1}{c_{44}} & 0 \\
            0 & 0 & 0 & 0 & 0 & \frac{1}{c_{44}} \\          
        \end{array} \right)
    \end{equation}
with $d=c_{11}^2+c_{11}c_{12}-2c_{12}^2$, and hence, using equation \ref{eqn:dualTensorDef} for tensor $\Omega _{kl}$, we find the partial relaxation volumes
    \begin{eqnarray}
        \label{eqn:partialRelaxationVolumes}
        \Omega^{(1)} = \Omega^{(2)} &=& \frac{a}{c_{11}+2 c_{12}} - \frac{b}{c_{44}}     \nonumber   \\
        \Omega^{(3)} &=& \frac{a}{c_{11}+2 c_{12}} + 2\frac{b}{c_{44}}.     
    \end{eqnarray}
We can define a single dimensionless measure of the anisotropy of relaxation for the structures considered here as a ratio of the smallest to largest (magnitude) partial relaxation volumes, 
    \begin{equation}
        \label{eqn:anisotropy}
        \alpha \equiv \frac{\Omega^{(1)}}{\Omega^{(3)}}.
    \end{equation}
The value $\alpha=1$ indicates that all the three partial relaxation volumes are equal, as should be the case for a spherical void. A value in the interval $0<\alpha<1$ indicates that the principal lattice relaxation is along the Burgers vector, but there are also smaller relaxations in the two orthogonal directions with the same sign (i.e. compressive or tensile ) as the principal relaxation. The value $\alpha=0$ indicates that the only lattice relaxation is along the Burgers vector. A value $\alpha<0$ indicates that the principal lattice relaxation is along the Burgers vector, but there are also smaller relaxations in the two orthogonal directions with the opposite sign (i.e. compressive or tensile ) as the principal relaxation.

For a $\half\langle 111 \rangle$ loop, the anisotropy parameter is
    \begin{equation}
        \label{anisotropy_111}
        \alpha_{\half\langle 111 \rangle} = \frac{ c_{44} a - ( c_{11} + 2c_{12} ) b }{ c_{44} a + 2 ( c_{11} + 2c_{12} ) b }.
    \end{equation}
Given the value of $\alpha$ and the total relaxation volume $\Omega_{rel}$ we can reconstruct the dipole tensor for a $\half[ 111 ]$ loop as 
    \begin{eqnarray}
        \label{eqn:reconstructDipoleTensor111}
        P_{11}=P_{22}=P_{33}=a&=&\frac{c_{11}+2c_{12}}{3}\Omega _{rel}                 \nonumber \\
        P_{12}=P_{23}=P_{31}=b&=&\frac{c_{44} ( \alpha - 1 )}{ 3(1 + 2 \alpha )} \Omega _{rel}      
    \end{eqnarray}

The same process can be followed for a $[ 001 ]$ loop, for which the dipole tensor is 
    \begin{equation}
        P_{[001]} = \left( \begin{array}{ccc} a & 0 & 0 \\ 0 & a & 0 \\ 0 & 0 & a' \end{array} \right).
    \end{equation}    
This gives partial relaxation volumes
    \begin{eqnarray}
        \Omega^{(1)} = \Omega^{(2)} &=& \frac{c_{11} a }{d} - \frac{ c_{12} a'}{d}     \nonumber   \\
        \Omega^{(3)} &=& -\frac{ 2 c_{12} a}{d} + \frac{( c_{11} + c_{12} ) a'}{d},
    \end{eqnarray}
and the anisotropy coefficient
    \begin{equation}
        \label{anisotropy_100}
        \alpha_{\langle 100 \rangle} = \frac{ c_{11} a - c_{12} a' }{ - 2 c_{12} a + ( c_{11} + c_{12} ) a' }.
    \end{equation}
The parameter $d$ is defined above.    
Given $\alpha$ and the total relaxation volume $\Omega _{rel}$, we can also reconstruct the dipole tensor for a $[001]$ loop as 
    \begin{eqnarray}
        P_{11}=P_{22}=a&=&d\frac{ c_{12}(\alpha-1) + c_{11}\alpha  }{ ( c_{11}^2 + c_{11}c_{12} + 2 c_{12}^2 ) ( 1 + 2 \alpha) } \Omega _{rel}        \nonumber \\
        P_{33}=a'&=&d\frac{ c_{11} + 2 c_{12}\alpha  }{ ( c_{11}^2 + c_{11}c_{12} + 2 c_{12}^2 ) ( 1 + 2 \alpha) } \Omega _{rel}     \nonumber \\
        P_{12}=P_{23}=P_{31}&=&0  
    \end{eqnarray}        
    
For spherical voids the coefficient $\alpha_{\mbox{sphere}} = 1$.
The dipole tensor for a spherical void can be reconstructed as \cite{Dudarev_PRM2018} 
    \begin{eqnarray}
        \label{eqn:reconstructDipoleTensorSphere}
        P_{11}=P_{22}=P_{33} &=& \frac{c_{11}+2c_{12}}{3}\Omega _{rel}  = K\Omega _{rel}                 \nonumber \\
        P_{12}=P_{23}=P_{31} &=& 0,
    \end{eqnarray}        
where $K$ is the bulk modulus.    

\begin{figure}[h!tb!]
\centering
\includegraphics[width=7.5cm]{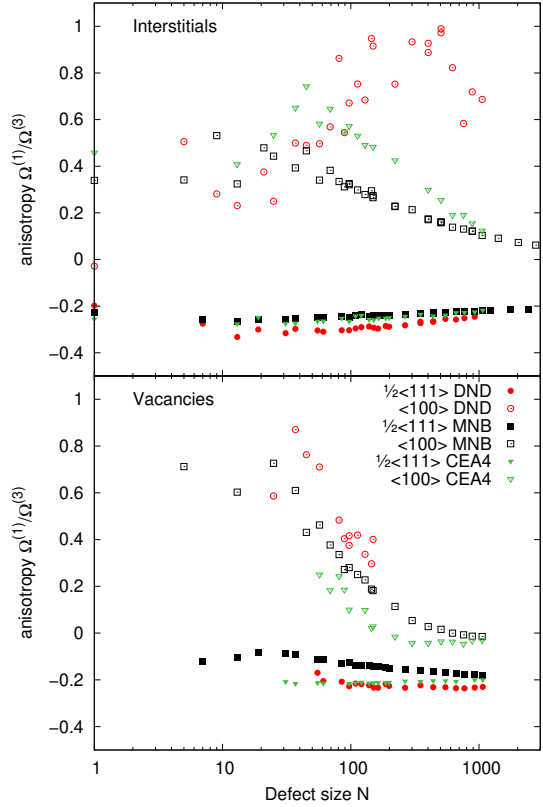}\\
\caption{
The relaxation volume anisotropy $\alpha$, as defined by the ratio of the smallest to the largest partial relaxation volume (see equation \ref{eqn:anisotropy}), for idealised circular prismatic loops.
} 
\label{fig:anisotropy}       
\end{figure}

We choose to present a parameterization required to fully reconstruct the dipole tensor as the pair of values $\{ \Omega_{rel},\alpha \}$.
The latter is plotted in figure \ref{fig:anisotropy}.
Fits to this data are given in table \ref{tab:anisotropy}.

\begingroup
\begin{table*}
\centering
\begin{tabular}{   l | l | rrrr   }
method 	&	structure   &   $\alpha_0$     &   $\alpha_1$  &   $\alpha_2$    &    $\alpha_3$   \\
\hline
MNB		&   i$\langle 111\rangle$   & -0.195 & $	-0.144	\pm 	0.001	$ & $ 	0.189	\pm 	0.006	$ & $	-0.212	\pm 	0.018	$ \\ 
MNB		&   i$\langle 100\rangle$   & 0      & $	0.196	\pm 	0.007	$ & $ 	2.320	\pm 	0.050	$ & $	-2.470	\pm 	0.139	$ \\ 
MNB		&   v$\langle 111\rangle$   & -0.195 & $	0.020	\pm 	0.003	$ & $ 	0.514	\pm 	0.017	$ & $	-0.331	\pm 	0.044	$ \\          
MNB		&   v$\langle 100\rangle$   & 0      & $	0.567	\pm 	0.041	$ & $ 	-3.158	\pm 	0.250	$ & $	19.777	\pm 	0.662	$ \\ 
\hline
DND		&   i$\langle 111\rangle$   & -0.203 & $	-0.255	\pm 	0.004	$ & $ 	0.270	\pm 	0.030	$ & $	-0.216	\pm 	0.089	$ \\ 
DND		&   i$\langle 100\rangle$   &        & $	5.826	\pm 	0.075	$ & $ 	-19.512	\pm 	0.469	$ & $	17.645	\pm 	1.081	$ \\ 
DND		&   v$\langle 111\rangle$   & -0.203 & $	-0.226	\pm 	0.004	$ & $ 	0.557	\pm 	0.029	$ & $	3.666	\pm 	0.112	$ \\ 
DND		&   v$\langle 100\rangle$   & 0      & $	-0.254	\pm 	0.064	$ & $ 	5.499	\pm 	0.361	$ & $	-0.907	\pm 	1.250	$ \\ 
\hline
CEA4	&   i$\langle 111\rangle$   & -0.199 & $	-0.128	\pm 	0.004	$ & $ 	0.036	\pm 	0.028	$ & $	-0.069	\pm 	0.081	$ \\
CEA4	&   i$\langle 100\rangle$   & 0      & $	0.157	\pm 	0.033	$ & $ 	4.574	\pm 	0.212	$ & $	-4.630	\pm 	0.563	$ \\
CEA4	&   v$\langle 111\rangle$   & -0.199 & $	-0.028	\pm 	0.002	$ & $ 	0.103	\pm 	0.012	$ & $	-0.961	\pm 	0.030	$ \\
CEA4	&   v$\langle 100\rangle$   & 0      & $	0.091	\pm 	0.023	$ & $ 	-3.012	\pm 	0.156	$ & $	36.838	\pm 	0.510	$ \\
\end{tabular}
\caption{
The dimensionless relaxation volume anisotropy parameter $\alpha$, as defined by the ratio of the smallest to the largest partial relaxation volume ( see equation \ref{eqn:anisotropy} ), for idealised circular prismatic loops, fitted to $\alpha = \alpha_0 + \alpha_1 \ln{ N }/\sqrt{N} + \alpha_2 /\sqrt{N} + \alpha_3/N$. 
The constant term $\alpha_0$ is derived using linear elasticity (equations \ref{anisotropy_111} and \ref{anisotropy_111}), using the computed values for elastic constants ( see table \ref{tab:elastic_constants} ).
}
\label{tab:anisotropy}
\end{table*}
\endgroup

\subsection{Cascade simulations}
\label{Cascades}

The final set of configurations we consider are taken from the MD cascade simulations dataset provided for the IAEA Visualisation Challenge \cite{IAEAVisChallenge}, generated according to the methodology detailed in Refs. \cite{Sand_EPL2013,Sand_MRL2017}. The simulations were evolved using a stiffened DND potential \cite{Bjo09} until a simulated time 40ps. The simulation cell was initially a perfect crystal, with damping applied to atoms with kinetic energy over 10 eV, and with an additional thermostat at the boundaries of the supercell \cite{Berendsen_JCP1998}. One atom was given an initial energy of 50-150keV, representing a primary knock-on event, and the final temperature was under 1K.
The cascade configurations were relaxed at constant volume using the procedure detailed in Ref. \cite{Dudarev_NF2018} with different EAM potentials, and the relaxation volume computed using the stress method.

\begin{figure}[h!tb!]
\centering
\includegraphics[width=7.5cm]{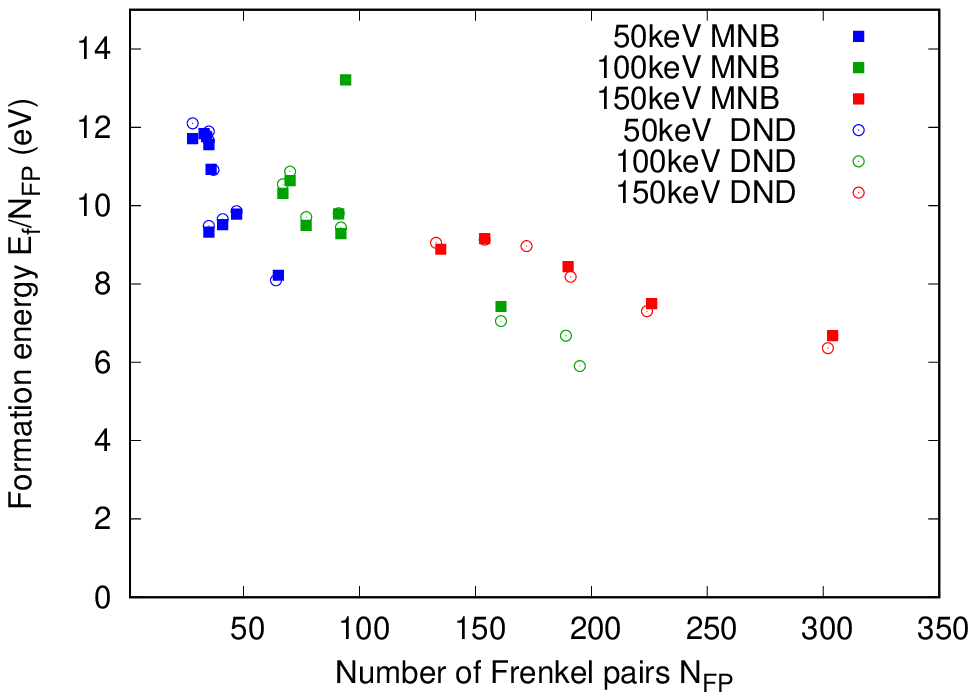}\\
\includegraphics[width=7.5cm]{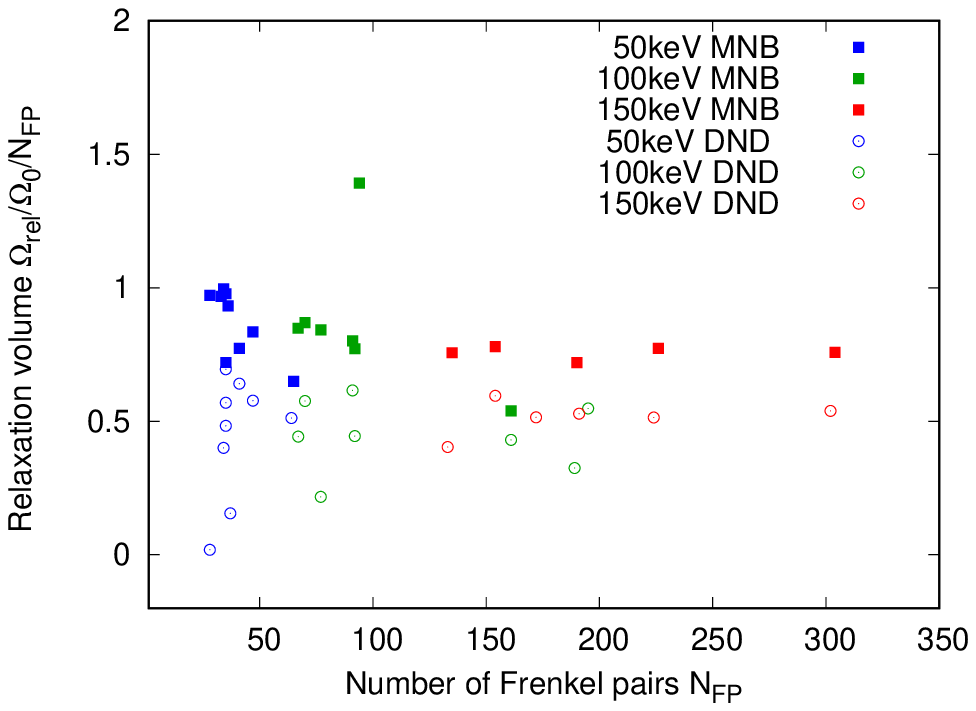}\\
\caption{
Formation energy and relaxation volumes of defect clusters generated by MD cascade simulations.
Filled symbols: computed with MNB potential, open circles with DND potential.
The formation energy of the cascades computed with the two potentials is very similar, but MNB tends to produce a larger relaxation volume.
Note that the cascade configurations were generated with the DND potential, then relaxed with both MNB and DND.
} 
\label{fig:cascades}       
\end{figure}

The results for the energy and relaxation volume as a function of the number of Frenkel pairs produced are shown in figure \ref{fig:cascades}.
We see that there is a slight tendency for a lower energy per Frenkel pair for the largest cascades.
This may indicate that the largest cascades ( in terms of Frenkel pairs produced ) are associated with the largest dislocation loops, and the largest loops have the lowest formation energy per point defect.
The relaxation volume shows a clear correlation with the Frenkel pair count, but for individual cascades there can be considerable variation.  
This is consistent with the preceding results, if the defect clusters are weakly interacting, as the relaxation volume should be determined by the degree of clustering rather than the total number of defects. On the basis of these simulations, it is reasonable to give a single relaxation volume per cascade.  We find from the relaxed cascades
	\begin{equation}    
	    \label{eqn:volume_per_cascade}
		\Omega _{rel}/\Omega _0 = b_0 N_{FP},
    \end{equation}
with $b_0 = 0.77 \pm 0.01$ for MNB and $b_0 = 0.50 \pm 0.02$ for DND.

\begin{figure}[h!tb!]
\centering
\includegraphics[width=7.5cm]{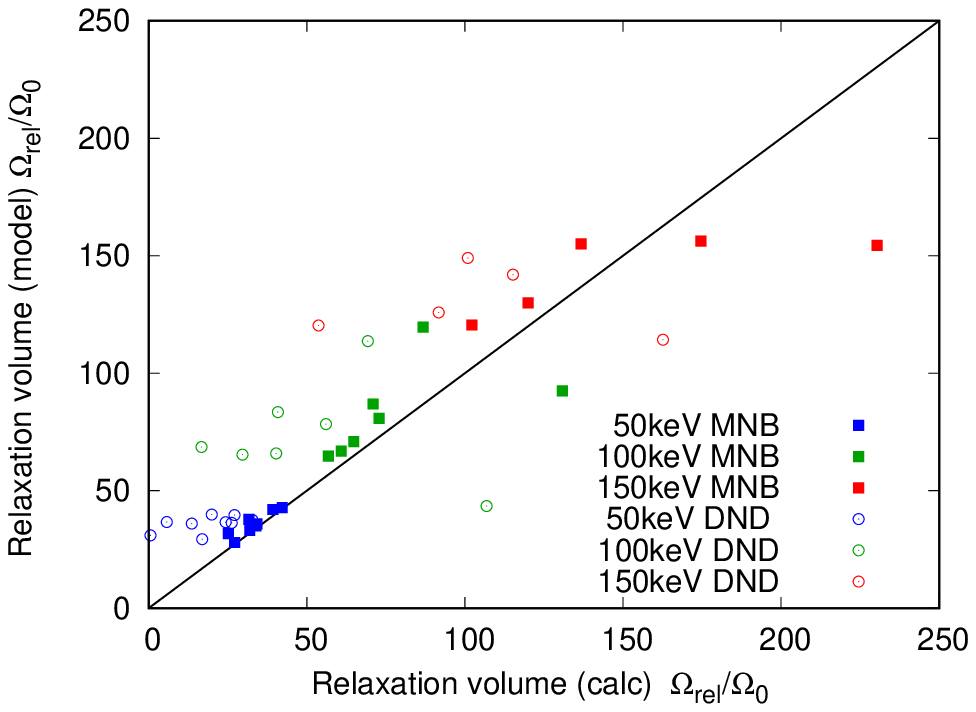}\\
\caption{
A comparison of the relaxation volume computed using a full relaxation of the cascade (x-axis) to the relaxation volume predicted using the tabulated fits to the data ( tables \ref{tab:volume_lowest},\ref{tab:regression_volume_loop},\ref{tab:regression_energyvolume_void} ).
The diagonal line indicates a 1:1 match- ie a perfect reproduction of the relaxation volume.
} 
\label{fig:cascade_fit}       
\end{figure}

Finally we can establish the predictive quality of the tabulated data presented here by using tables \ref{tab:volume_lowest},\ref{tab:regression_volume_loop},\ref{tab:regression_energyvolume_void} to estimate the relaxation volume of a cascade configuration.
After the initial Wigner-Seitz analysis of the cascade configuration, clusters of interstitials and vacancies were grouped where pairs of like-character point defects were separated by nearest- or next-nearest neighbours.
As many clusters are too small to perform a DXA analysis \cite{Stukowski_MSMSE2012}, we assert that all the interstitial loops are of $\half\langle 111\rangle$ type.
The relaxation volume for each defect is then looked up, and the total volume summed.
The result is displayed in figure \ref{fig:cascade_fit}.
While there is some scatter for MNB potential, and possibly additional relaxation within the cascade for the DND potential, it is clear that the simple empirical fits for mesoscale defect relaxation volume give a reasonable estimate of the cascade relaxation volume.
This is an important result for the transferability of our approach.
It indicates that even in the extreme case of high-energy defects formed close together in a cascade, the interaction between defects has a small effect on their relaxation volumes, and so elastically at least the defects can be treated as quasi-independent.
This therefore demonstrates that the relaxation volume is a good phase field for multiscale modelling.

\section{Comparison with isotropic linear elasticity}
\label{comparison_linear_elasticity}

In section \ref{anisotropy} we defined the anisotropy parameter in the partial relaxation volumes of idealised loop defects considered in section \ref{LowEnergyClusters}. The expected values of the anisotropy parameter can be computed using linear elasticity as follows.
The dipole tensor for a dislocation loop with normal $\hat{\mathbf{n}}$ and Burgers vector $\mathbf{b}$ and area A in isotropic elasticity is given by \cite{Dudarev_Acta2017}
    \begin{equation}
        P_{ij} =  A \left[ \mu \left( b_i n_j + b_j n_i \right) + \lambda b_k n_k \delta_{ij} \right],
        \label{Pij_loop}
    \end{equation}
where $\mu=(c_{11}-c_{12})/2$ and $\lambda=c_{12}$ are the shear modulus and Lam\'e constant.
Einstein summation over repeated indices is assumed.
If the normal and Burgers vector are parallel, as is the case here, this expression simplifies to
    \begin{equation}
        P_{ij} =  \pm N \Omega _0\left[ 2 \mu n_i n_j  + \lambda \delta_{ij} \right],
    \end{equation}
where the sign is positive for the interstitial loop and negative for a vacancy loop. For the $\half[ 111 ]$ loop this reduces further to 
    \begin{eqnarray}
        P_{11}=P_{22}=P_{33}&=& \pm K N \Omega _0               \nonumber \\
        P_{12}=P_{23}=P_{31}&=& \pm \frac{2 \mu}{3} N \Omega _0  
    \end{eqnarray}       
and so, substituting into equation \ref{eqn:partialRelaxationVolumes}, we can identify the relaxation volume for a $\half\langle 111 \rangle$ loop in isotropic linear elasticity as $\Omega _ {rel} = \pm N \Omega _0$ and its anisotropy as $\alpha= (-c_{11}+c_{12}+c_{44}) / (2c_{11} - 2c_{12} + c_{44}) = -1/5$. 
This negative value is significant, as it indicates that for a $\half\langle 111 \rangle$ interstitial loop the lattice expansion is negative in directions orthogonal to the Burgers vector.
Values of $\alpha$ computed using computed elastic constants are given in table \ref{tab:anisotropy}. Note that all the potentials give $\alpha\sim -0.2$. 

Similarly, for a $\half[ 001 ]$ loop we find $P_{11}=P_{22}= \pm \lambda N \Omega _0$, $P_{33}=\pm (2\mu+\lambda) N \Omega _0$, and $P_{12}=P_{23}=P_{31}=0$,
and so the relaxation volume computed for a $\langle 100 \rangle$ loop in isotropic linear elasticity is also $\Omega _{rel} = \pm N \Omega _0$ but its anisotropy is $\alpha=0$.

\subsection{Relaxation volume of a vacancy cluster}
\label{relax_vol_void}

In this subsection, we derive an analytical formula for the relaxation volume of a mesoscopic spherical vacancy cluster, treating the problem in the linear elasticity approximation. The approach that we adopt here broadly follows the analysis by Wolfer and Ashkin  \cite{Wolfer_JAP1975}.   

The elastic displacement field around a spherical vacancy cluster, taken in the isotropic elasticity approximation, is
\begin{equation}
{\bf u}({\bf r})=\frac{C}{r^2}{\bf n} + D r {\bf n} ,\label{eqn:displacement}    
\end{equation}
where $C$ and $D$ are constant factors that we will derive from boundary conditions, and ${\bf n}={\bf r}/r$. The strain and stress fields associated with the spherical vacancy cluster are \cite{Dudarev_NF2018,Wolfer_JAP1975}
\begin{eqnarray}
\epsilon _{ij}({\bf r})=\frac{C}{r^2}\left(\delta_{ij}-3n_in_j\right) + D \delta_{ij},\nonumber \\
\sigma _{ij}({\bf r})=2\mu \frac{C}{r^2} \left(\delta_{ij}-3n_in_j\right) + 3K D \delta_{ij},\label{strain_stress}
\end{eqnarray}
where, as above, $\mu$ is the shear modulus of the material, and $K$ is the bulk modulus.

We now find the relaxation volume of a spherical vacancy cluster of radius $a$ embedded in a concentric spherical isotropic elastic medium of radius $R$.
Boundary conditions for surface tractions at the surface of the vacancy cluster ($r=a$) and at the outer surface ($r=R$) have the form
\begin{eqnarray}
    \left. \sigma _{ij}n_j \right|_{r=a} &= \left. -4\mu \frac{C}{r^3} n_i + 3 K D n_i\right|_{r=a} &= - p_a + \frac{2 \gamma}{a} , \nonumber   \\
    \left. \sigma _{ij}n_j \right|_{r=R} &= \left.-4\mu \frac{C}{r^3} n_i + 3 K D  n_i\right|_{r=R} &= - p_R - \frac{2\gamma}{R} , \nonumber\\  
\end{eqnarray}
where $p_a$ is the pressure of gas accumulated inside the vacancy cluster (for example helium or hydrogen), $p_R$ the external pressure, and $\gamma$ is the orientation-average surface energy. 
From the boundary conditions we find
\begin{eqnarray}
    \label{eqn:relaxation_coefficients}
    C &=& \frac{ a^3 R^3 \left( \left(p_a - \frac{2 \gamma}{a}\right) - \left(p_R + \frac{2 \gamma}{R} \right)\right)  }{4 \mu \left( R^3 - a^3 \right)} \nonumber\\
    D &=&  \frac{  a^3 \left( p_a - \frac{2 \gamma}{a} \right) - R^3 \left( p_R + \frac{2 \gamma}{R} \right) }{3 K \left( R^3 - a^3 \right)}, 
\end{eqnarray}
which in the limit of zero external pressure ($p_R=0$) and $R\gg a$ simplify to 
\begin{eqnarray}
    \lim_{p_R=0,R\gg a} C &=& \frac{ a^3 }{4 \mu} \left( p_a - \frac{2 \gamma}{a} \right) \nonumber\\
   \lim_{p_R=0,R\gg a} D &=&  0. \nonumber 
\end{eqnarray}

Substituting the coefficients (equation \ref{eqn:relaxation_coefficients}) back into equation \ref{eqn:displacement} gives the magnitude of the displacement at the outer surface $u(R)$, and hence the change in volume. 
    \begin{equation*}
        \Omega_{rel} = \frac{4 \pi}{3} (R + u(R))^3 - \frac{4 \pi}{3} R^3      
    \end{equation*}        
After some rearrangement we find
    \begin{eqnarray}
        \Omega_{rel} &=&  \frac{ -4 \pi R^3}{3 (3K)^3} \left( p_R +\frac{2 \gamma}{R} \right)      \nonumber   \\
            && \times \left( \left( p_R +\frac{2 \gamma}{R} \right)^2 - 9 K \left( p_R + \frac{2 \gamma}{R} \right) + 27 K^2 \right)  \nonumber \\
        &-& \frac{ \pi a^3}{\mu (3K)^3} \left( \left(p_R + \frac{2 \gamma}{R}\right) - \left(p_a - \frac{2 \gamma}{a} \right) \right) \nonumber\\
        && \times
        \left( p_R + \frac{2 \gamma}{R} - 3K \right)^2 \left( 3K + 4 \mu \right)   
    \end{eqnarray}
The first term is the response of the outer surface to the external pressure and its surface energy, independent of the presence of the void in the interior.
The second term is the relaxation volume due to the void,
    \begin{eqnarray}
        \Omega_{rel} &=& - \frac{ \pi a^3}{\mu (3K)^3} \left( \left(p_R + \frac{2 \gamma}{R}\right) - \left(p_a - \frac{2 \gamma}{a} \right) \right) \nonumber  \\
        &&\times \left( p_R + \frac{2 \gamma}{R} - 3K \right)^2 \left( 3K + 4 \mu \right)
    \end{eqnarray}
In the limits $p_R=0$ and $R\gg a$, this simplifies to 
    \begin{eqnarray}
        \lim_{p_R=0,R\gg a} \Omega_{rel} &=& \frac{ \pi a^3 \left( p_a -  2 \gamma / a \right) ( 3 K + 4 \mu )}{ 3 K \mu }   \nonumber \\
        &=& \frac{3 \pi a^3 }{\mu}\left( \frac{ 1 - \nu }{ 1 + \nu }\right) \left( p_a - \frac{2 \gamma}{a} \right),
    \end{eqnarray}
where $\nu=c_{12}/(c_{11}+c_{12}) = \lambda/(2 \lambda + 2 \mu)$ is Poisson's ratio.

If the internal pressure in the vacancy cluster is zero ($p_a=0$), or in other words if there in no helium or hydrogen gas inside the vacancy cluster, the relaxation volume is negative and is proportional to the surface area of the cluster
\begin{equation}
    \Omega_{rel} \simeq - 6\pi \left(\frac{1-\nu}{1+\nu} \right)\frac{\gamma a^2}{\mu}.
\end{equation}
The radius of a spherical void can be related to the number of vacancies $N$ it contains, 
    \begin{equation}
        N \Omega _0 = \frac{4 \pi a^3}{3} , \nonumber
    \end{equation}
where $\Omega _0$ is the volume per atom, so we could also write
\begin{equation}
    \label{eqn:model_relax_vol}
    \Omega_{rel} \simeq - \left( \frac{243 \pi }{8} \right)^{1/3} \, \left(\frac{1-\nu}{1+\nu} \right)\frac{\gamma \mathrm{a}_0^2}{\mu} \, N^{2/3},
\end{equation}
where $\mathrm{a}_0$ is the lattice constant and $\Omega _0=\mathrm {a}_0^3/2$ for a bcc metal.

From this equation it follows that the relaxation volume {\it per vacancy} in a vacancy cluster (a void containing no gas) is negative and varies as the inverse third power of the number of vacancies that it contains
\begin{equation}
    \Omega_{rel}/N \sim - N^{-1/3},
\end{equation}
and vanishes in the macroscopic limit $N \gg 1$. 

We can compare the relaxation volumes computed using atomistic relaxations to the predictions from the above surface energy model, by substituting the elastic constants and spherically-averaged surface energies $\gamma$, as computed in section \ref{average_surface_energy} and tabulated in \ref{tab:elastic_constants}.
If we use the same surface energy model for the formation energy, we find 
    \begin{eqnarray}    
        \label{eqn:model_formation_energy}
        E_f &\simeq& 4 \pi \gamma a^2       \nonumber\\
            &\simeq& \left( 9 \pi \right)^{1/3} \, \gamma \mathrm{a}_0^2 \, N^{2/3}       \nonumber\\
    \end{eqnarray}
The comparison is shown in table \ref{tab:comparison_sim_model}. 
We conclude that the relaxation volume and formation energy of voids in tungsten are well reproduced by a simple surface energy model.    

\begingroup
\begin{table*}
\centering
\begin{tabular}{   l | cc | cc }
& \multicolumn{2}{c|}{$\Omega_{rel}/\Omega_0/N^{2/3}$}
 & \multicolumn{2}{c}{ $E_f/N^{2/3}$} \\
& calc & model & calc & model \\
\hline
DFT		&   -0.40   &   -0.42   &   6.66    &   7.08        \\
MNB		&   -0.50   &   -0.38   &   7.35    &   7.14        \\
DND		&   -0.31   &   -0.28   &   5.25    &   5.31        \\
CEA4	&   -0.122  &   -0.31   &   5.56    &   5.72        \\
\end{tabular}
\caption{
The relaxation volumes and formation energies (eV) in the limit of a large spherical void found by fitting to simulation data (calc), as tabulated in \ref{tab:regression_energyvolume_void} and using a surface energy only model (equations \ref{eqn:model_relax_vol} and \ref{eqn:model_formation_energy}).
}
\label{tab:comparison_sim_model}
\end{table*}
\endgroup

\section{Discussion of results}
\label{discussion}

\begin{figure}[h!tb!]
\centering
\includegraphics[width=7.5cm]{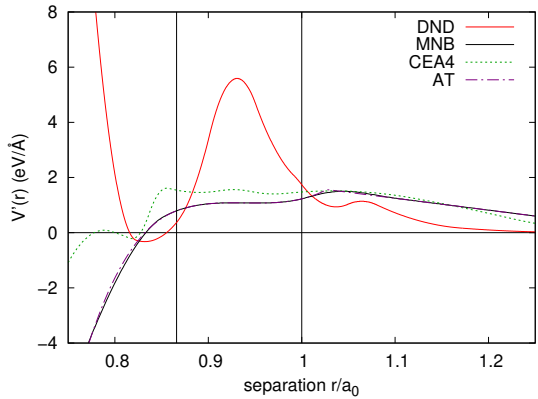}\\
\caption{
The first derivative of the pairwise part of selected empirical potentials, in the effective gauge \cite{Finnis_PMA1984,Bonny_JPCM2014} where $\rho_{eq}=1,F[0]=F[1]=0$.
The MNB potential is descended from the smooth Ackland-Thetford form \cite{Finnis_PMA1984,Ackland_PMA1987}, whereas the DND potential is a piecewise cubic spline which did not consider the first derivative during fitting.
Vertical lines are drawn at first and second nearest neighbour positions.
Note that the DND potential is not unstable at short separation, as might be inferred from this plot, as it is stabilised by its many-body part. In MD simulations the ZBL correction is also generally applied at short range\cite{SRIM,Bjo09}.
The Ackland-Thetford pairwise potential is very similar to the MNB.
} 
\label{fig:compare_EAM_dV}       
\end{figure}

It is instructive to consider why two potentials which give very similar energies of formation for lattice defects nevertheless give quite different elastic properties.
As shown in equation \ref{analyticDipoleTensorEAM}, the dipole tensor depends on the first derivative of the potential, and for the pairwise part $V(r_{ab})$ at least this is straightforward to analyse.
The MNB potential descends from the smooth Ackland-Thetford form\cite{Finnis_PMA1984,Ackland_PMA1987}, whereas the DND potential is based on a piecewise cubic spline form.
This latter form has a continuous second derivative, but discontinuous third, leading to a jagged second derivative. In the case of the DND potential, this second derivative swings from large positive to negative values. This in turn means that small changes in relative atom positions leads to large changes in the forces on the atoms.
The MNB potential, by contrast, has a fairly flat first derivative. 
The CEA-4 potential \cite{Marinica_JPCM2013}, which is also a cubic-spline form, but included fitting to forces during its construction, shares this flat first derivative for near equilibrium atom separations, but with more structure for greatly distorted structures.
This is illustrated in figure \ref{fig:compare_EAM_dV}.

In section \ref{Cascades} we found that the relaxation volume per cascade is proportional to the number of Frenkel pairs contained.
We can make an estimate for the relaxation volume per cascade if we make the assumption that the lattice defects are isolated and idealised.
Sand et al \cite{Sand_EPL2013} suggest that clusters should be produced in cascades with a frequency given by the power-law
    \begin{equation}
        f(N) = A/N^S,
    \end{equation}
and in ref \cite{Sand_EPL2016} give exponents for large clusters for interstitial-type and vacancy clusters in bulk tungsten cascade simulations as $S_I=1.6$ and $S_V=2.0$ respectively.
The expected relaxation volume per Frenkel pair is then
	\begin{eqnarray}    
	    \label{eqn:volume_per_cascade_powerLaw}
		\Omega _{rel}/\Omega _0/N_{FP} &=& \sum_N q_I(N) \Omega _I(N) +  q_V(N) \Omega_V(N)  \nonumber\\
    \end{eqnarray}    
with $q_{I/V}(N)$ being the weighting for interstitial/vacancy clusters of size $N$,
    \begin{equation}
        q_I(N) = \frac{ N^{-S_I}}{ \sum_N N^{1-S_I} }  \quad , \quad q_V(N) = \frac{ N^{-S_V}}{\sum_N N^{1-S_V}},
    \end{equation}
and $\Omega_{I/V}(N)$ given by the fits in tables \ref{tab:regression_energyvolume_void}, \ref{tab:volume_lowest}, and \ref{tab:regression_volume_loop}.
This gives the values 
    $\Omega_{rel}/\Omega _0= 0.87 N_{FP}$ for MNB. For DND and CEA4 the prefactor is $0.80$ and $1.31$ respectively.
Though not a perfect match, this simple calculation returns that MNB has a larger relaxation volume than DND, and that both have a scaling factor a little under unity.
This is observed in the expensive atomic relaxations in figure \ref{fig:cascades}.

\section{Conclusions}

In this paper we have computed the relaxation volumes of a wide range of lattice defects in tungsten, using three empirical potentials and density functional theory.
We have presented the data in a number of scatter plots, but have also presented tabulated data for empirical fits to the results.
It is hoped that the data in this form is readily applicable to a range of coarse grained models which include require either the elastic interactions between defects or the engineering stresses and strains induced by the defect population.

We have found that there is some considerable variation in absolute values of the relaxation volumes of defects compared across the EAM potentials we have considered.
This is because the defect elastic properties were never considered in the parameter fitting.
We were, however, able to identify some cross-potential trends.
The relaxation volume per point defect varies according the specific configuration of the defect cluster, but for small defect clusters ($N<10$) is likely to remain in the range $\pm 10\%$. 
This means that the size and character ( vacancy- or interstitial- type ) is sufficient information to predict the elastic properties.
Larger clusters ($N>10$) are most stable as dislocation loops, categorised by a Burgers vector, or in the case of vacancies are more stable as voids ($N<6\times 10^5$ for MNB or $N<3\times 10^6$ for CEA4\cite{Fikar_NME2018}, and we find $N<8\times 10^6$ for DND). 
$\half \langle 111 \rangle$ and $\langle 100 \rangle$ dislocation loops and voids are sufficiently dissimilar to require their own branches to predict their elastic properties.

We showed in section \ref{Cascades} that the relaxation volume of a cascade is proportional to the number of Frenkel pairs it contains, with a positive coefficient around unity. The structure in the relaxation volume per point defect is to some extent averaged out by the range of sizes of defects produced in a cascade.
The expected number of Frenkel pairs per cascade is itself proportional to the PKA energy, according to the NRT formula \cite{Norgett_NED1975}, a result broadly confirmed by MD simulation \cite{Calder_JNM1993,Becquart_JNM2000}.
Hence we can say that as a rule-of-thumb, the relaxation volume per cascade increases linearly with PKA energy.
As the defected microstructure evolves, annihilation between vacancy- type and interstitial- type will reduce the relaxation volume significantly, preserving the linearity with Frenkel pair count. 
Coalescence of small defect clusters will have a smaller effect on the relaxation volume. 
As we have found a potential-dependent non-monotonic variation of the relaxation volume of individual dislocation loops with defect count $N$, it is not clear at this point whether coalescence will increase or decrease the relaxation volume. 
DFT calculations of the dipole tensors of large loops may be able to answer this question in the future.

\section{Acknowledgements}
This work has been carried out within the framework of the EUROfusion Consortium and has received funding from the Euratom research and training programme 2014-2018 under grant agreement No 633053 and from the RCUK Energy Programme [grant number EP/P012450/1]. The views and opinions expressed herein do not necessarily reflect those of the European Commission. DNM acknowledges the support from high-performace computing facility MARCONI
(Italy) through the EUROfusion HPC project AMPSTANI (2017-2018).

\section*{References}

\bibliographystyle{unsrt}

\appendix
\section{Average surface energy}
\label{average_surface_energy}
In this appendix we find the average surface energy suitable for a spherical void given calculated values of the surface energy on facets.
As we are working with cubic crystals, we expand the surface energy in cubic harmonics
    \begin{equation}
        Y_{mn}(x,y,z) = \left( x^4 + y^4 + z^4 \right)^m  \left( x^2 y^2 z^2 \right)^n,
    \end{equation}
where $x,y,z$ are direction cosines, so that the surface energy at a general direction is interpolated as
    \begin{equation}
        g(x,y,z) = \sum_{mn} a_{mn} Y_{mn}(x,y,z)
    \end{equation}
We can fit available surface data to the lowest orders of $Y_{mn}$.
It is most common in the literature to see data for $\langle 110 \rangle$, $\langle 100 \rangle$, $\langle 211 \rangle$, and $\langle 111 \rangle$ planes, in which case it suffices to take $0 \le m,n \le 1$.
Writing the surface energy for the $\langle 110 \rangle$ plane as $\gamma_{\langle 110 \rangle}$, and similarly for the others, we find
    \begin{eqnarray}
        a_{00} &=& -\gamma_{\langle 100 \rangle} + 2\gamma_{\langle 110 \rangle}       \nonumber\\
        a_{10} &=& 2( \gamma_{\langle 100 \rangle} - \gamma_{\langle 110 \rangle})       \nonumber\\
        a_{01} &=& 27( \gamma_{\langle 100 \rangle} + 3 \gamma_{\langle 111 \rangle} - 4\gamma_{\langle 211 \rangle} )       \nonumber\\
        a_{11} &=& -54( \gamma_{\langle 100 \rangle} + 3 \gamma_{\langle 111 \rangle} - 6\gamma_{\langle 211 \rangle} + 2 \gamma_{\langle 110 \rangle} ). 
    \end{eqnarray}
The spherically-averaged surface energy,$\gamma$, can be found by integrating over the surface of the sphere:
    \begin{eqnarray}
        \label{eqn:average_surface_energy}
        \gamma &=& \frac{1}{4\pi} \, \int_{\theta=0}^{\pi} \int_{\phi=0}^{2 \pi}  g( \sin \theta \cos \phi, \sin \theta \sin \phi, \cos \theta) \sin \theta d\theta d\phi       \nonumber\\
        &=& a_{00} + \frac{a_{01}}{105} + \frac{3 a_{10}}{5} + \frac{a_{11}}{231}       \nonumber\\
        &=& \frac{1}{385} \left( 86 \gamma_{\langle 100 \rangle} + 128 \gamma_{\langle 110 \rangle} + 27 \gamma_{\langle 111 \rangle} + 144 \gamma_{\langle 211 \rangle} \right).        \nonumber\\
    \end{eqnarray}

It should be noted that two assumptions are made here, firstly that voids are unfacetted, and secondly that the surface energy varies smoothly with the orientation of the facet, so this interpolation should not be applied uncritically to other cases.
However, this simple expression does give a single value for surface energy suitable for use in our analytical calculations of void relaxation volumes.

\end{document}